\documentclass{ar2e-arxiv}

\usepackage{graphics}
\usepackage{psfrag}
\usepackage{hyperref}

\begin{document}

\vspace*{-72pt}
\hbox to \textwidth{%
         \hss{\vbox to 3cm{%
             \hbox{MZ-TH/10-07}
             \hbox{CERN-PH-TH/2010-068}
             \hbox{KEK-PREPRINT-2010-9}\vss
}}}
\vspace{-24pt}

\input symbols.sty
\input results.sty

\newcommand{\arXiv}[2]{}

%
\renewcommand{\arXiv}[2]{, \hbox{\href{http://arxiv.org/abs/#1}{arXiv:#1}#2}}

\makeatletter
\def\myep@[#1]#2{\resizebox{#1\textwidth}{!}{\includegraphics{#2}}}
\def\myep@c[#1]#2{\centerline{\myep@[#1]{#2}}}
\def\myeps{\@ifnextchar[{\myep@}{\myep@[1]}}
\def\myepsc{\@ifnextchar[{\myep@c}{\myep@c[1]}}
\def\at{@}
\makeatother

\def\Scp{{\cal S}_{CP}}
\def\Acp{{\cal A}_{CP}}

\def\BAR#1{{\overline{#1}}{}}

\def\Bbar{\BAR B}
\def\Kbar{\BAR K}
\def\bbar{\BAR b}
\def\cbar{\BAR c}
\def\sbar{\BAR s}
\def\ubar{\BAR u}
\def\dbar{\BAR d}
\def\qbar{\BAR q}
\def\pbar{\BAR p}
\def\nubar{\BAR \nu}
\def\lbar{\BAR \ell}

\def\Tbar{\BAR T}
\def\mbar{\BAR m}

\def\BBbar{B\Bbar}
\def\qqbar{q\qbar}

\jname{}
\jyear{}
\jvol{}
\ARinfo{}  

\title{Radiative and Electroweak Penguin Decays of B Mesons}

\markboth{Tobias Hurth and Mikihiko Nakao}{Radiative and Electroweak Penguin Decays of B Mesons}

\author{Tobias Hurth
 \affiliation{Inst.  for Physics, Johannes Gutenberg University, D-55099 Mainz,
   Germany;  \,\, email: Tobias.Hurth@cern.ch}
 and Mikihiko Nakao
 \affiliation{KEK, High Energy Accelerator Research Organization,
   Tsukuba, 305-0801, Japan and the Graduate University for Advanced
   Studies (Sokendai), Tsukuba, 305-0801, Japan; \,\, email:
   mikihiko.nakao\at kek.jp} 
}


\begin{abstract}
The huge datasets collected at the two $B$ factories, Belle and BaBar,
have made it possible to explore the radiative penguin process
$\btosgamma$, the electroweak penguin process $\btosll$ and
the suppressed radiative process $\btodgamma$ in detail, all  in
exclusive channels and inclusive measurements.  Theoretical tools 
have  also advanced to meet or surpass the experimental precision,
especially in inclusive calculations and the various ratios of exclusive
channels.  In this article, we review the theoretical and experimental
progress over the past  decade  in the radiative and electroweak penguin 
decays of $B$ mesons.
\end{abstract}

\maketitle

                         \section{INTRODUCTION}

\begin{figure}[ht]
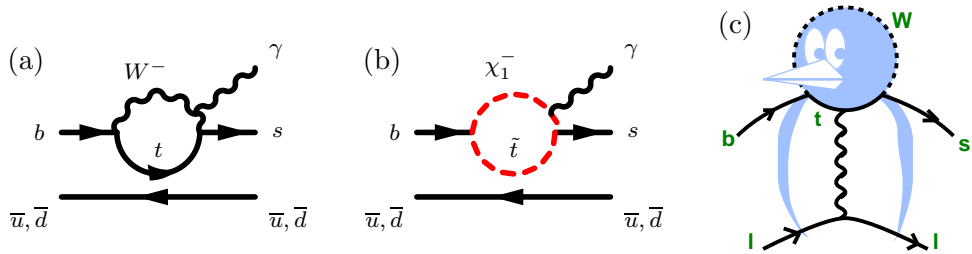

  \begin{center}
    \begin{minipage}{0.3\textwidth}
      (a)\\[-5mm]
      \myeps{theory/feynmp-b2xsgam}
    \end{minipage}%
    \hspace{0.05\textwidth}%
    \begin{minipage}{0.3\textwidth}
      (b)\\[-5mm]
      \myeps{theory/feynmp-b2xsgam_susy}
    \end{minipage}%
    \hspace{0.05\textwidth}%
    \begin{minipage}{0.25\textwidth}
      (c)\\[-5mm]
      \myeps{theory/nakao-penguin}
    \end{minipage}
  \end{center}
  \caption{Examples of radiative penguin decay diagrams
    (a) in the Standard Model and (b) beyond. (c) A penguin.}
  \label{fig:diagrams}
\end{figure}

The $B$ meson system, which is a bound state that consists of a $b$ quark and a light
antiquark, provides an ideal laboratory for precise  study of the
Standard Model (SM) of particle physics, and thus facilitates the search for
new physics (NP).  Because  the $b$ quark mass is much larger than
the typical scale of the strong interaction, the otherwise troublesome
long-distance strong interactions are generally less important and are
under better control than in other lighter meson systems.  Radiative
penguin%
\footnote{The name penguin decays was first introduced in
  Ref.~\cite{Ellis:1977uk} as the result of a bet.  A more detailed
  account of the name can be found in Ref.~\cite{Shifman:1995hc}.}
decays of the $B$ meson
with the emission of a photon ($\gamma$) and electroweak penguin decays
with the emission of a lepton pair ($\ell^+\ell^-$, $\ell=e,\,\mu$)
are of particular interest in this respect.
These processes reveal the flavor changing neutral current (FCNC), that
is the transition of a $b$ quark with an electric charge of $-1/3$ into an
$s$ or a $d$ quark of the same charge.  In the SM,
the FCNC occurs only via virtual loop diagrams
(Fig.~\ref{fig:diagrams}).
Additional NP contributions to these decay rates are not necessarily 
suppressed with respect to the SM contribution.  Examples of such NP 
scenarios include those in which the SM particles in the loop diagram 
are replaced by hypothetical new particles at a high mass scale; so far,
they have not been directly 
accessible in collider experiments.
Radiative and electroweak penguin decays are highly sensitive to NP 
because they are theoretically
well-understood and have been extensively measured
at the $B$ factories.  The search for such NP effects 
complements
the search for new particles produced at collider experiments.

The first generation of the $B$ factories at KEK (the Belle experiment at
the KEKB $\epem$ collider)~\cite{Belle} and at SLAC (the BaBar experiment at
the PEP-II $\epem$ collider)~\cite{Babar} have collected huge samples of
$B$ meson decays and have thereby established the SM picture of $CP$
violation and other flavor-changing processes in the quark sector.
These processes are
governed by a single $3\times3$ unitarity matrix referred to as the
Cabibbo-Kobayashi-Maskawa (CKM)
matrix~\cite{Kobayashi:1973fv,Cabibbo:1963yz}.  The  CKM matrix  can be illustrated
by a unitarity triangle in the complex plane that is overconstrained
by measurements from the $B$ factories, the
Tevatron $B$ physics programs (namely the CDF~\cite{TevatronB1} and
D0~\cite{TevatronB2} experiments), and earlier kaon decay experiments.
In other words, none of the current  measurements of $B$ meson decays have 
observed any unambiguous sign of
NP~\cite{Buchalla:2008jp,Antonelli:2009ws}.  
Although this experimental result is an impressive  success 
of the CKM theory within the SM, there is still
room for sizable new effects from new flavor structures, given that  FCNC
processes have been tested up to only the $10\%$ level.

The nonexistence of large NP effects in the FCNC processes hints at the
famous flavor problem, namely why FCNCs  are suppressed. 
This problem must be solved in any viable NP model.
Either the mass scale of the new degrees of freedom is very high or the 
new flavor-violating couplings are small for reasons that remain to be found. 
For example,  assuming  generic new flavor-violating couplings,  the present data on  
$K$-$\Kbar$ mixing implies  a very high NP scale of order $10^3$--$10^4$ TeV 
depending on whether the new  contributions enter at loop-level  or at tree-level.   
 In contrast, theoretical
considerations on the Higgs sector, which is responsible for the mass generation 
of the fundamental particles in the SM, call for NP at
order $1$ TeV.    As a consequence,
any NP below the 
$1$-TeV scale must have a nongeneric flavor structure.  
The present measurements of $B$ decays, especially of FCNC
processes, already significantly restrict the parameter
space of NP models. For further 
considerations on NP, the reader is referred to another article in this
volume~\cite{GinoYossi} and to Ref.~\cite{BurasFlavour}.

Quark-level FCNC processes such as $\btosgamma$, to which NP may
contribute, cannot be directly measured because  the strong interaction
forms hadrons from the underlying quarks.
Instead, the experimentally measured and theoretically calculated
process is a
$B$ meson decay into a photon plus an inclusive hadronic
final state $X_s$, which  includes all the hadron combinations that carry 
 the
strange quantum number $s=+1$ of the $s$ quark.%
\footnote{%
  In this review, we use  the following notations and conventions: We denote the 
  inclusive decay as $B\to X_s\gamma$
  when charge conjugation is implied, or as $\Bbar\to X_s\gamma$
  and $B\to X_{\sbar}\gamma$ to reflect the quark charges of the
  underlying processes  $\btosgamma$ and $\bbar\to\sbar\gamma$,
  respectively, when $CP$ and angular asymmetries are concerned.
  Here, $B$ denotes either an isospin- and $CP$-averaged state of $B^0$, $\Bbar^0$, $B^+$ and $B^-$ mesons, or an
  isospin averaged state of $B^0$ and $B^+$ (in the latter case, $\Bbar$
  denotes $\Bbar^0$ and $B^-$).
  Expressions
  are  constructed similarly for inclusive $X_d\gamma$ and
  $X_s\elel$ final states, and isospin-averaged exclusive final states.
  In the literature, the notation $\Bbar \to X_s\gamma$ is also commonly
  used for the case that includes  charge conjugation.}
Exclusive final states with one or a few specific hadrons in the final state (e.g., $B\to K^*\gamma$) 
have less predictive  power theoretically; however, because  the measurements are easier and
better  defined, there are other useful observables beyond branching
fractions, in particular $CP$, forward-backward, isospin,  and polarization
asymmetries.  In the future,  a  large overconstrained set of measurements of
these observables will allow us to detect specific patterns and to
distinguish between various NP scenarios.

This review covers progress in radiative and electroweak decays in the past decade, during 
which  a huge number of $B$ factory results were   accumulated and  significant progress 
in various theoretical aspects was  achieved.  
The pioneering work that led to  the first observation of the $\btosgamma$ process
by CLEO~\cite{cleo-kstgam1993,cleo-xsgam1995}  was  discussed in an
earlier volume of this journal~\cite{Lingel1998ar}.

Our  review is organized as follows.  In
Section~\ref{sec:th-framework}, we describe theoretical tools for
radiative and electroweak penguin decays, and in
Section~\ref{sec:ex-techniques} we describe experimental techniques.  We give 
theoretical predictions  in
Section~\ref{sec:th-predictions} and summarize the  measurements of radiative and
electroweak penguin decays in Section~\ref{sec:ex-results}.
Finally, we briefly discuss future prospects in Section~\ref{sec:outlook}.

                    \section{THEORETICAL FRAMEWORK}
                    \label{sec:th-framework}

Inclusive $B$ decays are theoretically clean because they are dominated by partonic 
(perturbatively calculable) contributions. Nonperturbative corrections are in general rather
small~\cite{Hurth:2003vb,Hurth:2003ej,Hurth:2007xa}. 
This result can be derived with the help of the heavy mass expansion  (HME)  of the inclusive decay rates 
in inverse powers of the $b$ quark mass. Up-to-date predictions of exclusive $B$ decays are based 
on the quantum chromodynamics (QCD)-improved factorization (QCDF) and soft collinear effective theory  (SCET) methods. In general, exclusive
modes have larger nonperturbative QCD corrections than do inclusive modes.

\subsection{Electroweak Effective Hamiltonian}
\label{electroweakhamiltonian}

Rare $B$ decays are governed by an interplay between the weak and strong
interactions.  The QCD
corrections that arise from hard gluon exchange bring in large logarithms of the
form $\alpha_s^n(m_b) \,  \log^m(m_b/M)$, where $M=m_t$ or $M=m_W$ and
$m \le n$ (with $n=0,1,2,...$).  These large logarithms are a natural feature in any
process in which  two different mass scales are present. To obtain  a
reasonable result, one must  resum at least the leading-log (LL)
series, $n=m$,  with the help of renormalization-group techniques.  Working to
next-to-leading-log (NLL) or  next-to-next-to-leading-log (NNLL) precision means that one 
resums  all the terms with   $n=m+1$ or  $n=m+2$. 
A suitable framework in which to achieve the necessary resummations of
the large logarithms is an effective low-energy theory with five quarks;
 this framework is obtained by integrating out the heavy particles, which in the SM are
the electroweak bosons and the top quark.

This  effective field theory approach serves as a theoretical framework
for both inclusive and exclusive modes.  The standard method of the
operator product expansion (OPE)~\cite{Wilson:1969zs,Wilson:1970ag} 
allows for a separation of the $B$
meson decay amplitude into two distinct parts:   the long-distance
contributions contained in the operator matrix elements and the
short-distance physics described by the so-called Wilson coefficients.

The electroweak effective Hamiltonian~\cite{Gaillard:1974nj,Altarelli:1974exa,Witten:1976kx}
can be written as
\begin{equation}
 {\cal H}_{\rm eff} =    \frac{4 G_{F}}{\sqrt{2}} \, 
\sum  {C_{i}(\mu, M)}\,\, \, {\cal O}_i(\mu), 
\end{equation}
where ${\cal O}_i(\mu)$ are the relevant operators and 
$C_{i}(\mu, M)$ are the corresponding Wilson coefficients.
As the heavy fields are integrated out, the complete top and
$W$ mass dependence is contained in the Wilson coefficients. 
Clearly, only within 
the observable ${\cal H}_{\rm eff}$  does the scale dependence cancel out.
$G_F$ denotes the Fermi coupling constant.

From the $\mu$ independence of the effective Hamiltonian, 
one can derive a renormalization group equation 
(RGE) for the Wilson 
coefficients $C_i(\mu)$:
\begin{equation}
\label{RGEa}
\mu \frac{d}{d\mu} C_i(\mu) = \gamma_{ji} \, C_j(\mu) ,
\end{equation}
where the matrix $\gamma$ is the anomalous dimension
matrix of the operators ${\cal O}_i$, which describes the 
anomalous scaling of the operators with respect to the 
 scaling  at the  classical level. 
At leading order, the solution is given by
\begin{equation}
\label{Wilsonsummation}
\tilde{C}_i(\mu)= \left[ 
\frac{\alpha_s (\mu_W)}{\alpha_s(\mu)} 
\right]^{\frac{\tilde{\gamma}^{0}_{ii}}{2 \beta_0}}  \, \tilde{C}_i (\mu_W) =
\left[\frac{1}{1+\beta_0\frac{\alpha_s(\mu)}{4\pi}\ln\frac{\mu^2_W}{\mu^2}}
\right]^{\frac{\tilde{\gamma}^{0}_{ii}}{2 \beta_0}} \, \tilde{C}_i (\mu_W) , 
\end{equation}
where  $\mu \, d/d\mu \, \alpha_s = -2 \beta_0 \alpha^2_s / (4 \pi)$, and
$\beta_0$ and $\tilde{\gamma}^0_{ii}$  correspond to  the leading anomalous 
dimensions  of the coupling constant and of  the operators, respectively. 
The tilde indicates that the diagonalized anomalous dimension matrix
is used.

Although  the Wilson coefficients $C_i (\mu)$ enter both inclusive and
exclusive processes and can be calculated with perturbative methods, the
calculational approaches to the matrix elements of the operators differ
  between the two  cases.  Within inclusive modes, one can use the quark-hadron duality  to derive a  well-defined HME  of the decay
rates in powers of $\Lambda/m_b$~\cite{Chay:1990da,Bigi:1992su,
  Bigi:1992ne,Bigi:1997fj,Manohar:1993qn,Manohar:2000dt}.  In
particular, the decay width of the $B \rightarrow X_s
\gamma$ is well approximated by the partonic decay rate, which can be
calculated in renormalization-group-improved perturbation theory~\cite{Falk:1993dh, Ali:1996bm}:
\begin{equation}
\Gamma ( B \rightarrow X_s \gamma) = \Gamma ( b \rightarrow X_s^{\rm
  parton} \gamma ) + \Delta^{\rm nonpert.}.
\end{equation}

In exclusive processes, however, one cannot rely on quark-hadron duality,
so one must estimate the matrix elements between
meson states.  A promising approach is the QCDF-method, which has been systematized for nonleptonic decays in the
heavy quark limit~\cite{Beneke:1999br,Beneke:2000ry,Beneke:2001ev}.
In addition, a more general quantum field theoretical framework for QCDF, known
as  SCET, has been proposed~\cite{Bauer:2000ew,Bauer:2000yr,Bauer:2001ct,Bauer:2001yt,Beneke:2002ph,Hill:2002vw}.
This method allows for a perturbative calculation of QCD corrections to
na\"{\i}ve factorization and is the basis for the up-to-date predictions for
exclusive rare $B$ decays. However, within this approach, a general
quantitative method to estimate the important $\Lambda/m_b$ corrections
to the heavy quark limit is missing.

\subsection{Perturbative Corrections to Inclusive Decays}
\label{perturbativecorrections}

Within inclusive $B$ decay modes, short-distance QCD effects are
very important.  For example, in the $B \rightarrow X_s
\gamma$ decay these effects  lead to a  rate enhancement by a factor of greater 
than two.  Such effects are induced by hard-gluon exchanges between the
quark lines of the one-loop electroweak diagrams.
The corresponding large logarithms have to be summed as discussed above.

The effective electroweak Hamiltonian that is relevant to $b  \to  s/d\,  \gamma$ and $b \to s/d\, \ell^+\ell^-$  transitions   in the SM reads 
\begin{equation}
 \label{Heff}
    {\cal H}_{\rm eff} =  - \frac{4G_F}{\sqrt{2}}
    \left[ \lambda_q^t \sum_{i=1}^{10} C_i {\cal O}_i  + 
           \lambda_q^u\sum_{i=1}^{2} C_i  ({\cal O}_i  - {\cal O}_i^u) \right]\, ,
\end{equation}
where the explicit CKM factors
are $\lambda_q^t = V_{tb}^{} V_{tq}^*$ and $\lambda_q^u = V_{ub}^{} V_{uq}^*$.
The unitarity relations $\lambda^{c}_q=-\lambda^{t}_q-\lambda^{u}_q$
have already been used. The dimension-six operators are 
\begin{equation} \begin{array}{rclrcl}
{\cal O}_1 &=& (\sbar_L\gamma_\mu T^a c_L)(\cbar_L\gamma^\mu T^a b_L)\,, &
{\cal  O}_2 &=& (\sbar_L\gamma_\mu c_L)(\cbar_L\gamma^\mu b_L)\,,\nonumber\\[1mm]
 {\cal O}_1^u &=& (\sbar_L\gamma_\mu T^a u_L)(\ubar_L\gamma^\mu T^a b_L)\,, &
{\cal  O}_2^u  &=& (\sbar_L\gamma_\mu u_L)(\ubar_L\gamma^\mu b_L)\,,\nonumber\\[1mm]
{\cal  O}_3 &=& (\sbar_L \gamma_\mu b_L){\sum}_q(\qbar\gamma^\mu q)\,, &
{\cal  O}_4 &=& (\sbar_L \gamma_\mu  T^a b_L){\sum}_q(\qbar \gamma^\mu  T^a q)\,,\nonumber\\[1mm]
{\cal  O}_5 &=& (\sbar_L  \Gamma  b_L){\sum}_q(\qbar \Gamma^\prime   q)\,, &
{\cal  O}_6 &=& (\sbar_L   \Gamma               T^a b_L){\sum}_q
(\qbar    \Gamma^\prime   T^a q)\,,\nonumber\\[1mm]
 {\cal O}_7    & = & {{e}\over{16\pi^2}} m_b (\sbar_{L} \sigma^{\mu\nu}
                b_{R}) F_{\mu\nu} \, , &
    {\cal O}_8    & = & {{g_s}\over{16 \pi^2}} m_b (\sbar_{L} \sigma^{\mu\nu}
                T^a b_{R}) G_{\mu\nu}^a \,, \nonumber \\[1mm]
    {\cal O}_9    & = & {{e^2}\over{16\pi^2}}(\sbar_L\gamma_{\mu} b_L)
                \sum_\ell(\lbar\gamma^{\mu}\ell) \,, &
    {\cal O}_{10} & = & {{e^2}\over{16\pi^2}}(\sbar_L\gamma_{\mu} b_L)
                \sum_\ell(\lbar\gamma^{\mu} \gamma_{5} \ell) \,, 
\end{array} \end{equation}
where $\Gamma = \gamma_\mu\gamma_\nu\gamma_\lambda$ and $\Gamma^\prime =
\gamma^\mu\gamma^\nu\gamma^\lambda$. The subscripts $L$ and $R$ refer to
left- and right-handed components, respectively,  of the fermion fields. In $b\to s$
transitions the contributions proportional to $\lambda_s^u$ are rather
small, whereas  in $b\to d$ decays,  where $\lambda_d^u$ is of the same order as
$\lambda_d^t$;  these  contributions  play an important role in $CP$ and isospin
asymmetries.  The semileptonic operators ${\cal O}_9$ and ${\cal
  O}_{10}$  occur only in the semileptonic $b \rightarrow s/d \, \ell^+
\ell^-$ modes.

Among the four-quark operators, only the
effective couplings for $i=1,2$ are large at the low scale $\mu = m_b$
[$C_{1,2}(m_b) \sim 1$],  whereas  the couplings of the other four-quark
operators have almost negligible values. But the dipole operators
[$C_7(m_b) \sim -0.3,\, C_8(m_b) \sim - 0.15$] and the semileptonic
operators [$C_9(m_b) \sim 4,\, C_{10}(m_b) \sim -4$] also play a
significant role.

There are three principal calculational steps that lead  to the LL (NNLL)
result within the effective field theory approach:
\begin{enumerate}
\item The full SM theory must  be matched with the effective theory at
  the scale $\mu=\mu_W$, where $\mu_W$ denotes a scale of order $m_W$ or
  $m_t$.  The Wilson coefficients $C_i(\mu_W)$ pick up only  small QCD
  corrections, which can be calculated within  fixed-order perturbation
  theory.  In the LL (NNLL) program, the matching has to be worked out
  at the $O(\alpha_s^0)$ [$O(\alpha_s^2)$] level.
\item The evolution of these Wilson coefficients from $\mu=\mu_W$
  down to $\mu = \mu_b$ must then  be performed with the help of the
  renormalization group, where $\mu_b$ is of the order of $m_b$.  As the
  matrix elements of the operators evaluated at the low scale $\mu_b$
  are free of large logarithms, the latter are contained in resummed
  form in the Wilson coefficients. For the LL (NNLL) calculation, this RGE
  step has to be performed  using the anomalous-dimension matrix up to order
  $\alpha_s^1$ ($\alpha_s^3$).
\item To LL (NNLL) precision, the corrections to the matrix elements of
  the operators $\langle s \gamma |{\cal O}_i (\mu)|b \rangle$ at the scale $\mu =
  \mu_b$ must  be calculated to order $\alpha_s^0$ ($\alpha_s^2$)
  precision.  The calculation  also includes bremsstrahlung corrections.
\end{enumerate}

\paragraph{\boldmath$B \to X_s \gamma$}

The error of the LL prediction of the {$B \to X_s \gamma$} branching
fraction~\cite{Ciuchini:1993ks,Ciuchini:1993vr,Cella:1994np,Misiak:1992bc}
is dominated by a large renormalization scale dependence at the $\pm
25\%$ level, which indicates the importance of the
NLL series.
By convention, the dependence on the renormalization scale $\mu_b$ is
obtained by the variation $m_b/2 < \mu_b < 2 m_b$.  The three
calculational steps of the NLL enterprise---Step
1~\cite{Adel:1993ah,Greub:1997hf}, Step 2~\cite{Chetyrkin:1996vx,Gambino:2003zm}, 
and Step 3~\cite{Ali:1990tj,Greub:1996tg,Pott:1995if,Buras:2001mq,Buras:2002tp}---have been performed by many
different groups and have been independently checked.  
The resulting NLL
prediction had a small dependence on the scale $\mu_b$ as well as on the
matching scale $\mu_0$ below $5\%$.  But as first observed in
Ref.~\cite{Gambino:2001ew}, there was a large charm mass-scheme
dependence because the charm loop vanishes at  the LL level and
the significant charm dependence begins only at the NLL level.  By varying
$m_c/m_b$ in the conservative range $0.18 \le m_c/m_b \le 0.31$, which
covers both the pole mass value (with its numerical error) and the
running mass value $\mbar_c(\mu_c)$ with $\mu_c \in [m_c,m_b]$, one
finds an uncertainty of almost $10\%$~\cite{Hurth:2003dk,Hurth:2003pn}.
This uncertainty is the dominant error 
in the NLL prediction.  The renormalization scheme for $m_c$ is an NNLL
issue, and a complete NNLL calculation reduces
this large uncertainty by at least a factor of two~\cite{Asatrian:2005pm}.
This finding motivated the NNLL calculation of the $B \to X_s \gamma$
branching fraction.

\begin{figure}[ht]
\begin{center}
\myeps[0.4]{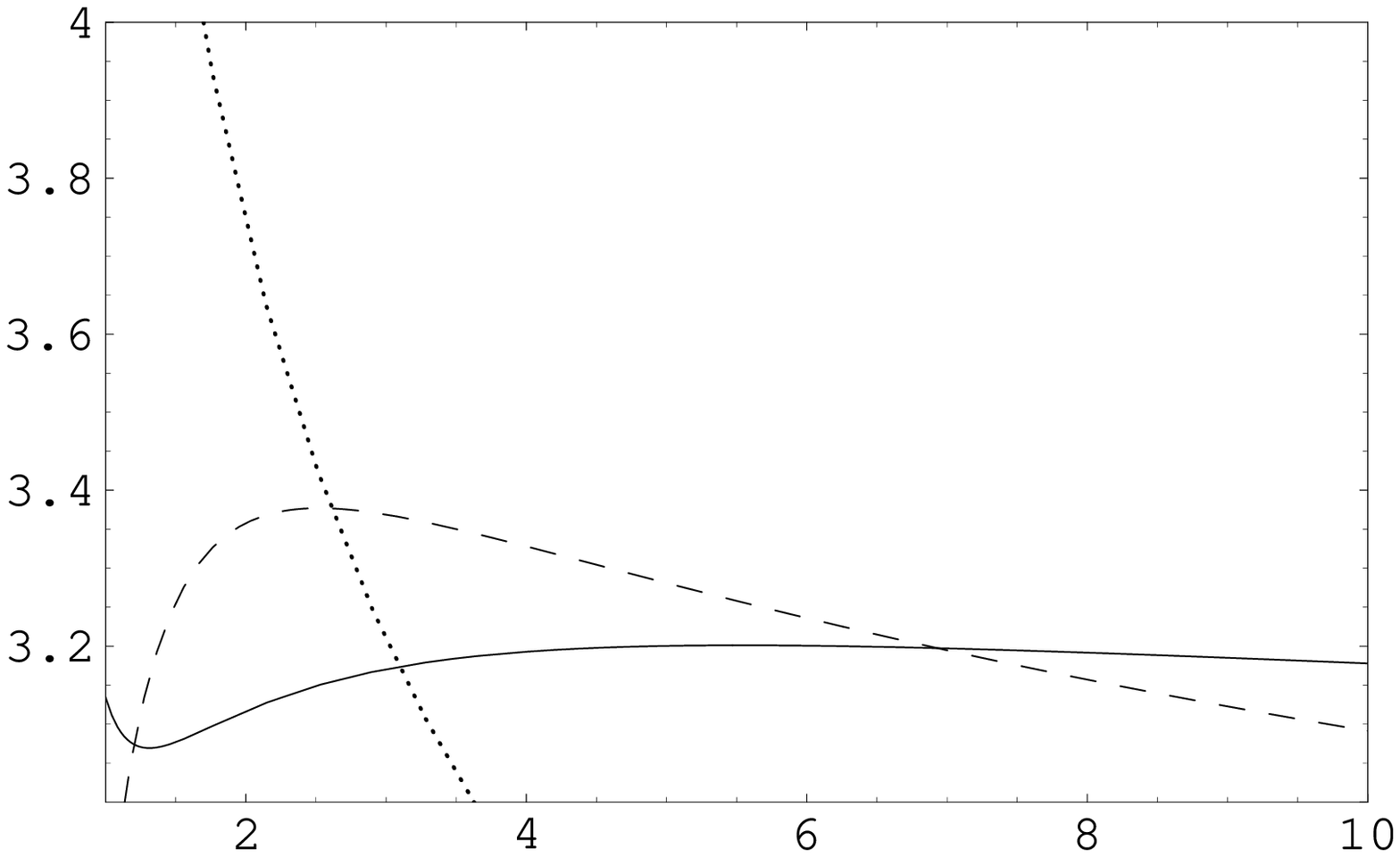}
\myeps[0.4]{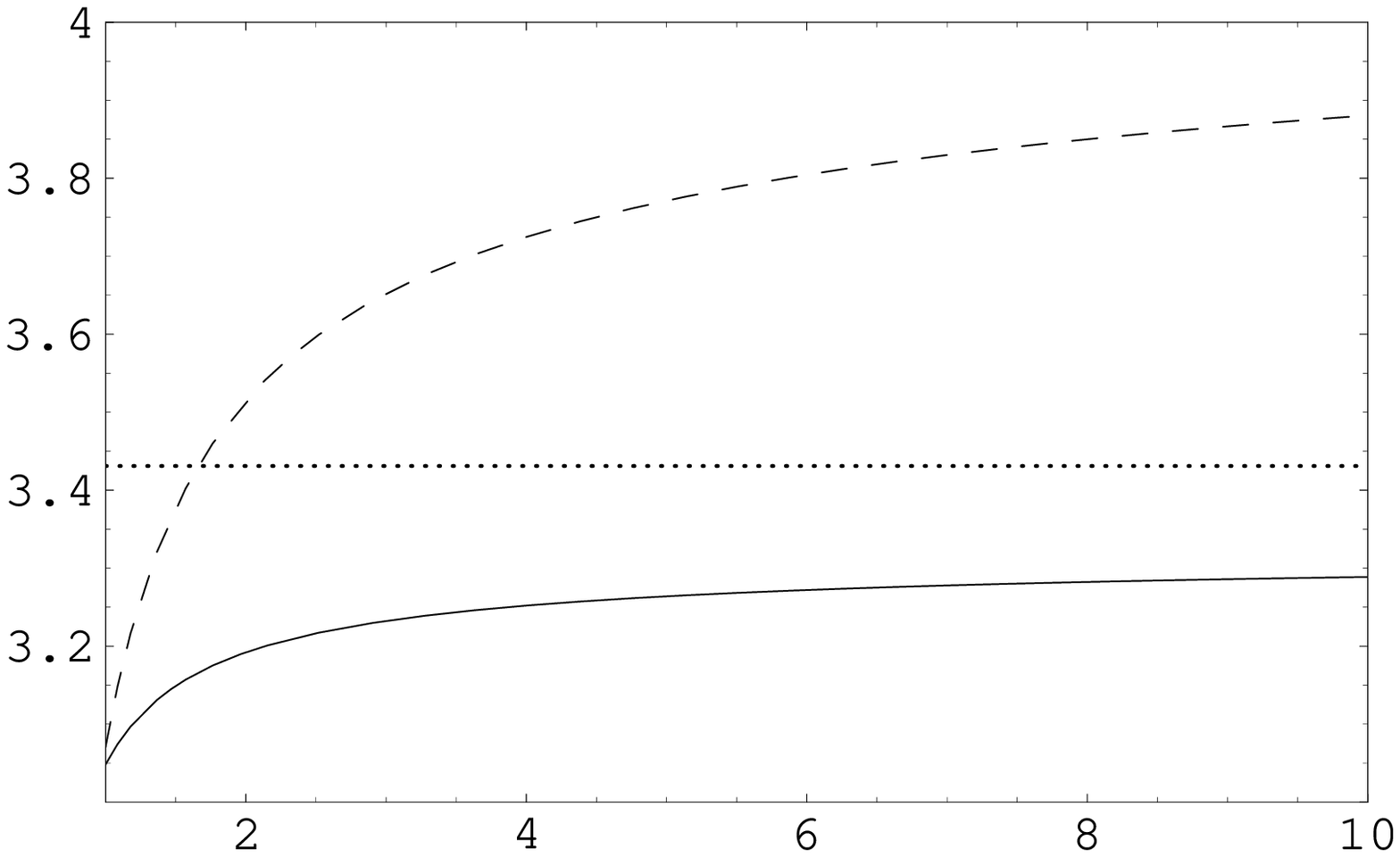}\\
$\mu_b\;$[GeV] \hspace{3.9cm}  $\mu_c\;$[GeV]
\caption{Renormalization-scale dependence of ${\cal B}(B \to X_s
  \gamma)$ in units $10^{-4}$ at leading log (dotted lines),
  next-to-leading log (dashed
  lines) and next-to-next-to-leading log (solid lines). The plots describe the dependence on
  (left) the the low-energy scale $\mu_b$ and (right) the charm mass renormalization
  scale $\mu_c$, from~\cite{Misiak:2006zs}. \label{fig:mudep}}
\end{center} 
\end{figure}

Following  a global effort, such an NNLL calculation was recently 
performed and led to the first NNLL prediction of the $B \to X_s
\gamma$ branching fraction~\cite{Misiak:2006zs}.  
This result is based on various highly-nontrivial perturbative
calculations ~\cite{Misiak:2004ew,Bobeth:1999mk,Gorbahn:2004my,Gorbahn:2005sa,Czakon:2006ss,Blokland:2005uk,Melnikov:2005bx,Asatrian:2006ph,Asatrian:2006sm,Bieri:2003ue,Misiak:2006ab}:
Within Step 1 the
matching of the effective couplings $C_i$ at the high-energy scale
$\mu_0 \sim M_W$ requires a three-loop calculation for the cases
$i=7,8$~\cite{Misiak:2004ew} and a two-loop calculation for the other
cases~\cite{Bobeth:1999mk}.  Within Step 2 the self-mixing of the
four-quark operators ($i=1,...,6$) and the self-mixing of the dipole
operators ($i=7,8$) have been  calculated by a three-loop calculation of
anomalous dimensions~\cite{Gorbahn:2004my,Gorbahn:2005sa}, and the
mixing of the four-quark operators into the dipole operators by a
four-loop calculation~\cite{Czakon:2006ss}.  These two steps have
established the effective couplings at the low scale $\mu_b \sim m_b$ to
NNLL precision.  Thus, large logarithms of the form $\alpha_s^{n+p}(m_b)
\, \log^n (m_b/m_W)$, $(p=0,1,2)$, are resummed.  Within Step 3, the
calculation of the matrix elements of the operators to NNLL precision,  
 only the dominating contributions have been
calculated or estimated by now.  The dominating two-loop matrix element
of the photonic dipole operator ${\cal O}_7$ including the
bremsstrahlung contributions has been calculated in
Refs.~\cite{Blokland:2005uk,Melnikov:2005bx,Asatrian:2006ph,Asatrian:2006sm}.
The other important piece is the three-loop matrix elements of the
four-quark operators, which has first been calculated within the
so-called large-$\beta_0$ approximation~\cite{Bieri:2003ue}.  A
calculation that goes beyond this approximation by employing an
interpolation in the charm quark mass $m_c$ from $m_c > m_b$ to the
\mbox{physical} $m_c$ value has been presented in
Ref.~\cite{Misiak:2006ab}. In this interpolation the $\alpha_s^2
\beta_0$ result~\cite{Bieri:2003ue} is assumed to be a good
approximation for the complete $\alpha_s^2$ result for vanishing charm
mass. It is this part of the NNLL calculation which is still open for
improvement.  Indeed a complete calculation of the three-loop matrix
elements of the four-quark operators ${\cal O}_{1,2}$ for vanishing
charm mass is work in progress~\cite{Boughezal:2007ny} and will
cross-check this assumption and the corresponding error estimate due to
the interpolation.

Some perturbative NNLL corrections have not yet been
included in the present NNLL estimate, but they are expected to be
smaller than the current perturbative uncertainty of $3\%$:
the virtual
and bremsstrahlung contributions to the $({\cal O}_7,{\cal O}_8)$ and
$({\cal O}_8,{\cal O}_8)$ interferences at order $\alpha_s^2$, the NNLL
bremsstrahlung contributions in the large-$\beta_0$-approximation beyond
the $({\cal O}_7,{\cal O}_7)$ interference term (which are already
available~\cite{Ligeti:1999ea}), the four-loop mixing of the four-quark
operators into the operator ${\cal O}_8$~\cite{Czakon:2006ss}, and the
exact mass dependence of various matrix elements beyond the large
$\beta_0$ approximation~\cite{Asatrian:2006rq, Boughezal:2007ny,
  Ewerth:2008nv}.

In the present NNLL prediction~\cite{Misiak:2006zs},  the reduction of the
renormalization-scale dependence at the NNLL is shown  in
Fig.~\ref{fig:mudep}.  The most important effect occurs for the charm
mass $\overline{\rm MS}$ renormalization scale $\mu_c$, which has been  the main
source of uncertainty at the NLL.  The current uncertainty of $\pm 3\%$
due to higher-order [${\cal O}(\alpha_s^3)$] effects can be estimated via 
the NNLL curves in Fig.~\ref{fig:mudep}. The reduction factor of the
perturbative error is greater  than a factor of three.  The central value of the
NNLL prediction is based on the choices $\mu_b=2.5$ GeV and $\mu_c=1.5$
GeV.

At NNLL QCD accuracy subdominant perturbative electroweak
two-loop corrections are also relevant and have 
been calculated to be $-3.6\%$~\cite{Czarnecki:1998tn,Kagan:1998ym,Baranowski:1999tq,Gambino:2000fz}.
They  are included in the present NNLL prediction.

\paragraph{\boldmath$B \rightarrow X_s \ell^+ \ell^-$}

Compared   with the $B \rightarrow X_s \gamma$ decay,
the inclusive $B \rightarrow X_s \ell^+ \ell^-$ decay presents a
complementary and more complex test of the SM, given that  different
perturbative electroweak contributions add to the decay rate. 
 This inclusive mode is also dominated by perturbative contributions, if
 one eliminates $c\cbar$ resonances with the help of kinematic cuts.
In the so-called perturbative $q^2$-windows below and above the
resonances, namely in the low-dilepton-mass region $1\;{\rm GeV}^2 < q^2
= m_{\ell\ell}^2 < 6\;{\rm GeV}^2$ as well as  in the high-dilepton-mass
region where  $q^2 > 14.4\;{\rm GeV}^2$, theoretical predictions for the
invariant mass spectrum are dominated by the perturbative contributions.
 A theoretical precision of order $10\%$ is possible.

Compared with the decay $B \rightarrow X_s \gamma$, the effective
Hamiltonian~(Eq.~\ref{Heff}) contains two additional operators
of $O(\alpha_{\rm em})$, the semileptonic operators ${\cal O}_9$ and ${\cal
  O}_{10}$.  Moreover, the first
large logarithm of the form $\log(m_b/m_W)$ already arises without
gluons, because the operator ${\cal O}_2$ mixes into ${\cal O}_9$ at one
loop.  It is then convenient to redefine the dipole and semileptonic
operators via
$\widetilde{{\cal O}}_i =  {16\pi^2} / {g_s^2} {{\cal O}}_i$, $\widetilde{C}_i = {g_s^2} / {(4\pi)^2} {C}_i$ for 
 $i=7,...,10$.
With this redefinition, one can follow the three calculational steps
discussed above.  In particular, after the reshufflings the one-loop
mixing of the operator ${\cal O}_2$ with $\widetilde{{\cal O}}_9$
appears formally at order $\alpha_s$. To LL precision, there is only
$\widetilde{{\cal O}}_9$ with a non-vanishing tree-level matrix element
and a non-vanishing coefficient.

It is well-known that this na\"{\i}ve $\alpha_s$ expansion is
problematic, since the formally-leading $O(1/\alpha_s)$ term in $C_9$ is
accidentally small and much closer in size to an $O(1)$ term.  Thus,
also specific higher order terms in the general
expansion are numerically important.%
\footnote{The $B \to X_s \ell^+\ell^-$
decay amplitude has the following structure ($\kappa=\alpha_{\rm em}/\alpha_s$): 
\begin{equation}
{\cal A}  =   \kappa \left[ {\cal A}_{LL}  + \alpha_s  \; {\cal A}_{NLL}+  
               \alpha_s^2 \; {\cal A}_{NNLL} + {\cal O}(\alpha_s^3) \right] \quad\quad  \mbox{with} \quad   {\cal A}_{LL} \sim  \alpha_s  \; {\cal  A}_{NLL}
\end{equation}
A strict NNLL calculation of the squared amplitude ${\cal A}^2$ should
only include terms up to order $\kappa^2 \alpha_s^2$.  However, in the
numerical calculation, one also includes the term ${\cal A}_{NLL} {\cal
  A}_{NNLL}$ of order $\kappa^2 \alpha_s^3$ which are numerically
important.  These terms beyond the formal NNLL level are proportional to
$|C_7|^2$ and $|C_8|^2$ and are scheme-independent. One can even argue
that one picks up the dominant NNNLL QCD corrections because the missing
NNNLL piece in the squared amplitude, namely ${\cal A}_{LL} {\cal  A}_{NNNLL}$, can safely be neglected~\cite{Huber:2008ak}.}

The complete NLL contributions have been  presented~\cite{Misiak:1992bc,Buras:1994dj}. For the NNLL calculation,  many
components  were taken over from the NLL calculation of the $B \to
X_s \gamma$ mode.  The additional components  for the NNLL QCD precision 
have been  calculated~\cite  {Bobeth:1999mk,Gambino:2003zm,Gorbahn:2004my,Asatryan:2001zw,Asatryan:2002iy,Ghinculov:2002pe,Asatrian:2002va,Asatrian:2003yk,Ghinculov:2003bx,Ghinculov:2003qd,Greub:2008cy,Bobeth:2003at,Huber:2008ak}:
Some new pieces for the matching to NNLL
precision (Step 1) have been  calculated in Ref.~\cite{Bobeth:1999mk}. To NNLL
precision the large matching scale uncertainty of $16\%$ at the NLL
level is  eliminated. In Step 2, the mixing of the four-quark
operators into the semileptonic operator ${\cal O}_9$ has  been
calculated~\cite{Gambino:2003zm,Gorbahn:2004my}. 
In Step 3, the four-quark matrix elements including
the corresponding bremsstrahlung contributions have been  calculated for the
low-$q^2$ region in
Refs.~\cite{Asatryan:2001zw,Asatryan:2002iy,Ghinculov:2003qd},
bremsstrahlung contribution for the forward-backward asymmetry in $B
\to X_s \ell^+\ell^-$ in Refs.~\cite{Ghinculov:2002pe,Asatrian:2002va,Asatrian:2003yk}, and the
four-quark matrix elements in the high-$q^2$ region in
Refs.~\cite{Ghinculov:2003bx,Ghinculov:2003qd,Greub:2008cy}.  The
two-loop matrix element of the operator ${\cal O}_9$ has been  estimated using
the corresponding result in the decay mode $B \to X_u \ell \nu$ and
also Pade approximation methods~\cite{Bobeth:2003at}; this estimate has been 
further improved in Ref.~\cite{Huber:2008ak}.

More recently  electromagnetic corrections were   calculated: NLL quantum
electrodynamics (QED)  two-loop corrections to the
Wilson coefficients are of $O(2\%)$~\cite{Bobeth:2003at}. 
Also, in the QED one-loop corrections to matrix elements, large
collinear logarithms of the form $\log(m_b^2/m^2_\ell)$ survive
integration if only a restricted part of the dilepton mass spectrum is
considered. These collinear logarithms add another contribution of order $+2\%$ in the low-$q^2$ region for $B \to X_s \mu^+\mu^-$~\cite{Huber:2005ig}. For  the high-$q^2$ region,  one finds 
$-8\%$~\cite{Huber:2007vv}.

\paragraph{\boldmath$B \to X_d \gamma$ and  $B \to X_d \ell^+ \ell^-$}

The perturbative QCD corrections in the inclusive decays $B
 \rightarrow X_ d \gamma$~\cite{Ali:1998rr,Hurth:2003dk,Hurth:2003pn}
and $B \to X_d \ell^+\ell^-$~\cite{Asatrian:2003vq,Seidel:2004jh}
can be treated completely  analogously  to those  in the corresponding $b
\rightarrow s$ modes.  The effective Hamiltonian is the same in these
processes,  up to the obvious replacement of the $s$ quark field by the
$d$ quark field.  However, because  $\lambda_u = V_{ub} V^*_{ud}$ for $b \to d
\gamma$ is not small with respect to $\lambda_t = V_{tb} V^*_{td}$ and
$\lambda_c = V_{cb} V^*_{cd}$, one must also account for the
operators proportional to $\lambda_u$, namely ${\cal O}^u_{1,2}$ in
Eq.~\ref{Heff}.  The matching conditions $C_i(m_W)$ and the solutions of
the RGEs, which yield  $C_i(\mu_b)$, coincide with those needed for
the corresponding $b \to s$ processes~\cite{Ali:1998rr,Asatrian:2003vq}.

\subsection{Hadronic Power Corrections to Inclusive Modes}
\label{hadronicpower}

The inclusive modes $B \rightarrow X_s \gamma$ and $B
\rightarrow X_s \ell^+ \ell^-$ are dominated by the partonic
contributions.  Indeed, if only the leading operator in the effective
Hamiltonian (${\cal O}_7$ for $B \to X_s \gamma$, ${\cal O}_9$ for $B \to X_s \ell^+\ell^-$) 
is considered, the HME makes it possible  to calculate
the inclusive decay rates of a hadron containing a heavy quark,
especially a $b$ quark~\cite{Chay:1990da,Bigi:1992su,
  Bigi:1992ne,Bigi:1997fj,Manohar:1993qn,Manohar:2000dt}.  The optical
theorem relates the inclusive decay rate of a hadron $H_b$ to the
imaginary part of the  forward scattering amplitude
\begin{equation}
\Gamma (H_b \rightarrow X) = \frac{1}{2 m_{H_b}} \Im \, \langle 
H_b \mid {{\bf T}} \mid H_b \rangle\, ,  
\end{equation}
where the transition operator ${\bf T}$  is given by 
${\bf T} = i \int d^4 x \,  T [ {\cal H}_{\rm eff} (x) {\cal H}_{\rm eff} (0)]$.
The insertion of a complete set of states, $| X\rangle\langle X |$,
leads to the standard formula for the decay rate:
\begin{equation}
\Gamma (H_b \rightarrow X) = 
\frac{1}{2 m_{H_b}} \sum_X (2 \pi)^4 \delta^4 ( p_i
- p_f) \mid \langle X \mid {\cal H}_{\rm eff} \mid H_b  \rangle \mid^2 \, . 
\end{equation}
It is then  possible to construct an 
OPE of  the operator ${\bf T}$, which 
is  expressed as a series of local operators that are  suppressed by powers of 
the $b$ quark mass and written in terms of the $b$ quark field:
\begin{equation}
T [ {\cal H}_{\rm eff}  {\cal H}_{\rm eff} ]  \stackrel{OPE}{=}  \frac{1}{m_b} \big( {\cal O}_0 + \frac{1}{m_b}
 {\cal O}_1 + \frac{1}{m_b^2} {\cal O}_2 + ... \big)\, . 
\end{equation}
This construction is based on the parton--hadron duality.
The sum is performed over all exclusive final states and that the
energy release in the decay is large with respect to the QCD scale,
$\Lambda \ll m_b$.  With the help of the heavy quark effective theory (HQET)~\cite{Isgur:1991xa,Neubert:1993mb}, namely the new heavy quark
spin-flavor symmetries that arise  in the heavy quark limit $m_b
\rightarrow \infty$, the hadronic matrix elements within the OPE,
$\langle H_b \mid {\cal O}_i \mid H_b \rangle$, can be further
simplified.  In this well-defined expansion, the free quark model is   
the first term in the constructed expansion in powers of
$1/m_b$ and,  therefore, is  the dominant contribution.  In the applications to
inclusive rare $B$ decays, one finds no correction of order $\Lambda/m_b$
to the free quark model approximation. The corrections to the
partonic decay rate begin  with $1/m_b^2$ only, which  implies
the rather small numerical impact of the nonperturbative corrections on 
the decay rate of inclusive modes. However,  there are more subtleties to consider 
if other than 
the leading operators are taken into account (see below). 

\paragraph{\boldmath$B \to X_s \gamma$}

These techniques can be used directly in the decay $B \rightarrow
X_s \gamma$ to single out nonperturbative corrections to the
branching fraction: If one neglects perturbative QCD corrections and
assumes that the decay $B \to X_s \gamma$ is due to the leading
electromagnetic dipole operator ${\cal O}_7$ alone, then the photon would
always be emitted directly from the hard process of the $b$ quark decay.
One has to consider the time-ordered product $T {\cal O}_7^+(x) \, {\cal
  O}_7(0)$.  Using the OPE for $T {\cal O}_7^+(x) \, {\cal O}_7(0)$ and
HQET methods, as discussed above, the decay width
$\Gamma(B \to X_s \gamma)$ reads~\cite{Falk:1993dh,Ali:1996bm} (modulo
higher terms in the $1/m_b$ expansion):
\begin{eqnarray}
\label{width}
\Gamma_{B \to X_s \gamma}^{({\cal O}_7,{\cal O}_7)} =
\frac{\alpha_{\rm em} G_F^2 m_b^5}{32 \pi^4} \, |V_{tb} V_{ts}|^2 \, C_7^2(m_b) \,
\left( 1 + \frac{\delta^{\rm NP}_{\rm rad}}{m_b^2} \right), \,\,\,
\delta^{\rm NP}_{\rm rad} = \frac{1}{2} \lambda_1 - \frac{9}{2} \lambda_2 ,
\end{eqnarray}
where $\lambda_1$ and $\lambda_2$ are the HQET parameters for the
kinetic energy and the chromomagnetic energy, respectively.  If the $B \rightarrow X_s
\gamma$ decay width is  normalized by the (charmless) semileptonic one, 
the nonperturbative corrections of order  $1/m^2_b$ cancel out within  
the ratio ${\cal B}(B \to X_s \gamma)/{\cal B}(B
\to X_u \ell \nu)$.

However, as noted in Ref.~\cite{Ligeti:1997tc}, there is no OPE for the
inclusive decay $B \rightarrow X_s \gamma$ if one considers
operators beyond the leading electromagnetic dipole operator ${\cal
  O}_7$.  Voloshin~\cite{Voloshin:1996gw} has identified a
contribution to the total decay rate in the interference of the $b \to s
\gamma$ amplitude due to the electromagnetic dipole operator ${\cal
  O}_7$ and the charming penguin amplitude due to the current-current
operator ${\cal O}_2$.
This resolved photon contribution contains subprocesses in which the photon couples to light partons
instead of connecting directly to the effective weak-interaction vertex~\cite{Benzke:2010js}.
If one treats  the charm quark as heavy, then it is possible
to expand the contribution in local operators. The
first term in this expansion may be the dominating one~\cite{Ligeti:1997tc,Grant:1997ec,Buchalla:1997ky}. This
nonperturbative correction is suppressed by $\lambda_2/m_c^2$ and is
estimated to be of order $3\%$ compared with the leading-order
(perturbative) contribution to the decay rate $\Gamma_{b \to s \gamma}$
which arises from the electromagnetic operator ${\cal O}_7$:
\begin{equation}
\label{voleffect}
\frac{\Delta \Gamma^{({\cal O}_2,{\cal O}_7)}_{B \to X_s \gamma
}}{\Gamma_{b \to s \gamma}^{\rm LL}} = -\frac{1}{9} \, \frac{C_2}{C_7} 
\frac{\lambda_2}{m_c^2} \simeq +0.03 .
\end{equation}
However, if the charm mass is assumed to scale as 
$m_c^2\sim\Lambda m_b$, then the charm penguin contribution must be
described by the matrix element of a nonlocal
operator~\cite{Ligeti:1997tc,Grant:1997ec,Buchalla:1997ky,Lee:2006wn}.

Recently, another example of such nonlocal matrix elements within
the power-suppressed contributions to the decay $B \to X_s \gamma$
 was  identified~\cite{Lee:2006wn}---specifically, in the interference of the
$b\to s\gamma$ transition amplitude mediated by the electro-magnetic
dipole operator ${\cal O}_7$, where  the $b \to s g $ amplitude is  mediated by
the chromo-magnetic dipole operator ${\cal O}_8$, followed by the
fragmentation of the gluon into an energetic photon and a soft
quark-antiquark pair.
A na\"{\i}ve dimensional estimate of these power corrections
leads to
\begin{equation}
\frac{\Delta \Gamma^{({\cal O}_7,{\cal O}_8)}_{B \to X_s \gamma
}}{\Gamma_{b \to s \gamma}^{\rm LL}}  \, \sim \, \pi\alpha_s  \ \frac{C_8}{C_7}\,  \frac{\Lambda}{m_b}\,,
\end{equation}
whereas an estimate using the vacuum insertion method for the nonlocal
matrix elements indicates an effect of $-3\%$.

Power corrections to the high-energy part of the $B \to X_s\gamma$
photon spectrum can be parameterized systematically in terms of
subleading shape functions.  For the interference of the ${\cal
  O}_7$--${\cal O}_7$ pair, these nonlocal operators reduce to local
operators, if one considers the total decay
rate~\cite{Lee:2004ja,Bosch:2004cb,Beneke:2004in}, whereas other resolved
photon contributions to the total decay rate---such as
the previously analyzed ${\cal O}_7$--${\cal O}_8$
interference term~\cite{Lee:2006wn}---cannot be described by a
local OPE.  A recent systematic analysis~\cite{Benzke:2010js} of
all resolved photon contributions related to other operators in the weak
Hamiltonian establishes this breakdown of the local OPE within the
hadronic power corrections as a generic result.
Clearly, estimating such nonlocal matrix elements is very difficult, and
an irreducible theoretical uncertainty of $\pm
(4-5)\%$ for the total $CP$ averaged decay rate, defined with a
photon-energy cut of
$E_\gamma = 1.6$ GeV, remains~\cite{Benzke:2010js}. This result strongly
indicates that the theoretical efforts for the $B \to X_s \gamma$
mode have reached the nonperturbative boundaries.
The complete effect of power corrections on $CP$ asymmetries has 
 not yet been estimated.

\paragraph{\boldmath$B \to X_s \ell^+ \ell^-$}

Hadronic power corrections in the decay {\bf $B \to X_s \ell^+
  \ell^-$} that scale  with $1/m_b^2$,
$1/m_b^3$~\cite{Falk:1993dh,Ali:1996bm,Chen:1997dj,Buchalla:1998mt,Bauer:1999kf,Ligeti:2007sn},
and $1/m_c^2$~\cite{Buchalla:1997ky} have also been considered.  They
can be calculated quite analogously to those in the decay $B
\rightarrow X_s \gamma$.  However, a systematic analysis of hadronic
power corrections including all relevant operators has yet to be performed.  Thus, an
additional uncertainty of $\pm 5\%$ should be added to all theoretical
predictions for this mode on the basis of  a simple dimensional estimate.

In the high-$q^2$ region, one encounters the breakdown of the HME 
 at the end point of the dilepton mass spectrum: Whereas the
partonic contribution vanishes, the $1/m_b^2$ and $1/m_b^3$
corrections tend towards a nonzero value. In contrast to the end-point
region of the photon-energy spectrum in the $B \rightarrow X_s
\gamma$ decay, no partial all-order resummation into a shape function is
possible. However, for an integrated high-$q^2$ spectrum an
effective expansion is found in inverse powers of $m_b^{\rm eff} = m_b
\times (1 - \sqrt{s_{\rm min}})$ rather than
$m_b$~\cite{Neubert:2000ch,Bauer:2001rc}. The expansion converges less
rapidly, depending on the lower dilepton-mass cut $s_{\rm min} =
q^2_{\rm min} / m_b^2$~\cite{Ghinculov:2003qd}.

The large
theoretical uncertainties could be significantly reduced by normalizing
the $B \rightarrow X_s \ell^+ \ell^-$ decay rate to the
semileptonic $B \rightarrow X_u \ell\nubar$ decay rate with the
same $q^2$ cut~\cite{Ligeti:2007sn}:
\begin{equation}
\label{eq:zoltanR}
{\cal R}(s_0) =  
\int_{\hat s_0}^1 {\rm d} \hat s \, {{\rm d} {\Gamma} (B \to X_s \ell^+\ell^-) \over {\rm d} \hat s}\,  /\,   
\int_{\hat s_0}^1 {\rm d} \hat s \, {{\rm d} {\Gamma} (B^0 \to X_u \ell \nu) \over {\rm d} \hat s}\,.
\end{equation}
For example,   the uncertainty due to the dominating $1/m_b^3$ term
would be 
reduced from $19\%$ to $9\%$~\cite{Huber:2005ig}. 

\paragraph{\boldmath$B \to X_d \gamma$}

The nonperturbative contributions in the decay $B \rightarrow X_d
\gamma$ can be treated analogously  to those in the decay $B
\rightarrow X_s \gamma$.  The  power corrections that scale as $1/m_b^2$
(in addition to the CKM factors) are the same for the two modes. Also, the systematic 
analysis of resolved contributions given in Ref.~\cite{Benzke:2010js} can be applied to  this case.
However, the long-distance contributions from the intermediate $u$ quark in
the penguin loops are critical.  They are suppressed in the $B
\rightarrow X_s \gamma$ mode by the CKM matrix elements. In
$B \rightarrow X_d \gamma$, there is no CKM suppression, and one must account for 
the nonperturbative contributions that arise from
the operator ${\cal O}_1^u$. 
The contribution 
due to the ${\cal O}_1^u$--${\cal O}_7$ interference scales with 
$\Lambda/m_b$~\cite{Buchalla:1998mt}. However, 
this interference  contribution 
vanishes in the total
$CP$-averaged  rate of $B \to X_s \gamma$ at order $\Lambda/m_b$~\cite{Benzke:2010js}. This result applies to 
the total rate of $B \to X_d \gamma$ as well. Other interference terms, namely the double resolved 
contributions ${\cal O}_1^u$--${\cal O}_8$ and ${\cal O}_1^u$--${\cal O}_1^u $, arise first at
order $1/m_b^2$, as they can also be deduced from the results presented in 
Ref.~\cite{Benzke:2010js}. Thus, there is no  power correction due to the operator ${\cal O}_1^u$ 
in the total rate of $B \to X_d \gamma$ at order $\Lambda/m_b$, which implies that the $CP$-averaged 
decay
rate of $B \rightarrow X_d \gamma$ is as theoretically clean as the decay rate 
of $B \rightarrow X_s \gamma$.

\paragraph{\boldmath$B \to X_d \ell^+ \ell^-$} In the case of {\bf $B \rightarrow X_d \ell^+
\ell^-$}
 long-distance contributions due to $u$ quark loops can be avoided
in the low-$q^2$ window $1\;{\rm GeV}^2 <q^2<6\;{\rm GeV}^2$.
The $\rho$ and $\omega$ resonances are below, and the $c\cbar$ ($J/\psi$,
$\psi^\prime$) resonances are above this
window~\cite{Asatrian:2003vq}. The effect of their respective tails
can be taken into account within the Kr\"uger-Sehgal (KS)
approach (see Section \ref{charmoniumresonance})~\cite{Kruger:1996cv,Kruger:1996dt}.  In this
low-$q^2$ region,  one can then treat the nonperturbative power
corrections analogously  to those  in the  decay $B
\rightarrow X_s \ell^+\ell^-$,  and one can expect a similar theoretical
accuracy in this $q^2$ window.

\subsection{Nonperturbative Corrections due to Kinematical Cuts}  
\label{nonperturbativecorrections}

There are additional subtleties in inclusive modes. Kinematical cuts induce additional
sensitivities to nonperturbative physics.

\paragraph{\boldmath$B \to X_s \gamma$}

\begin{figure}[ht]
\psfrag{x}[t]{$s$}
\psfrag{y}[b]{$d{\cal B}(B\to X_s\ell^+\ell^-)/ds\;\;[10^{-5}]$}
\centerline{
  \myeps[0.38]{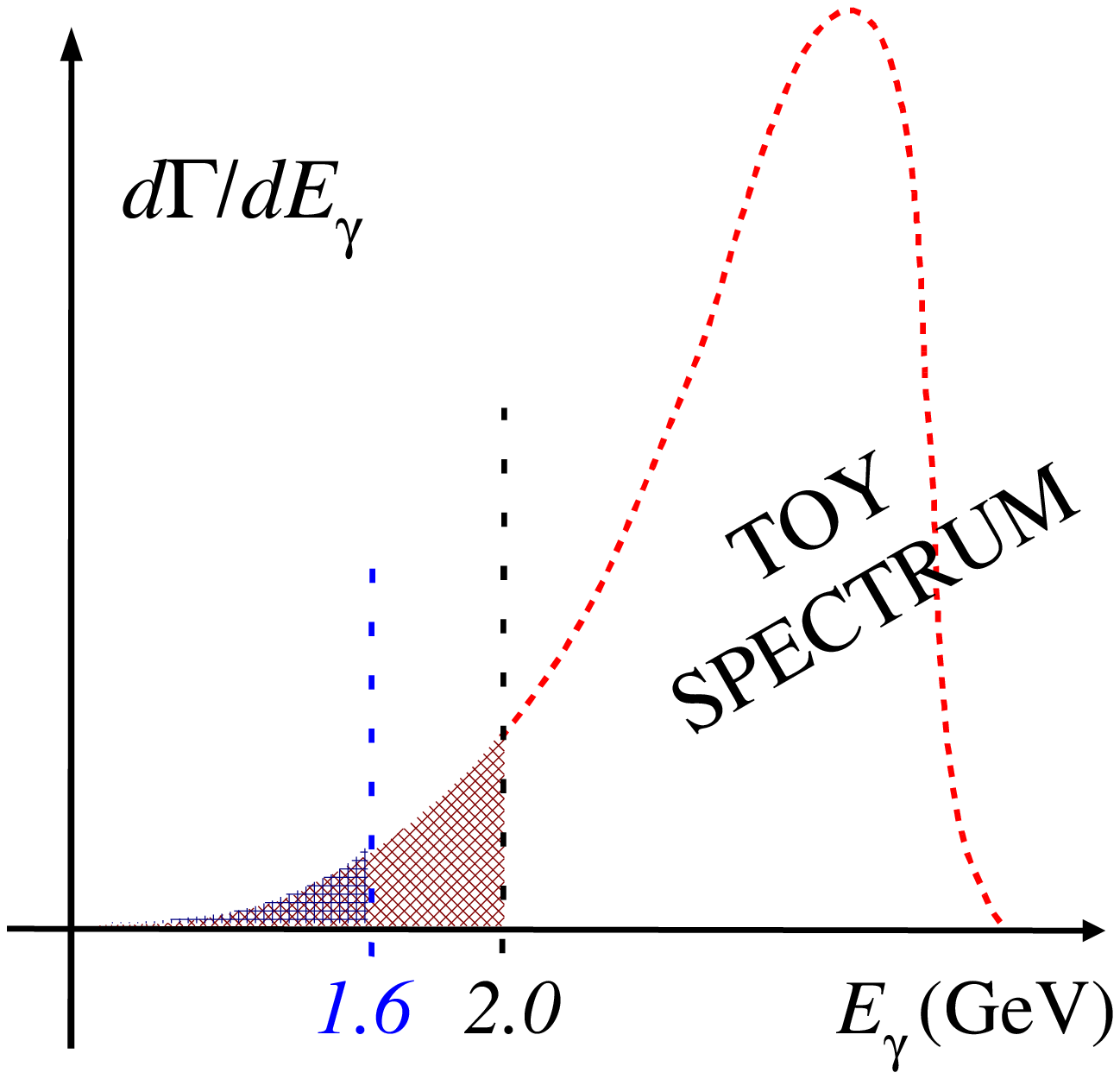}\hspace{0.05\textwidth}%
  \myeps[0.55]{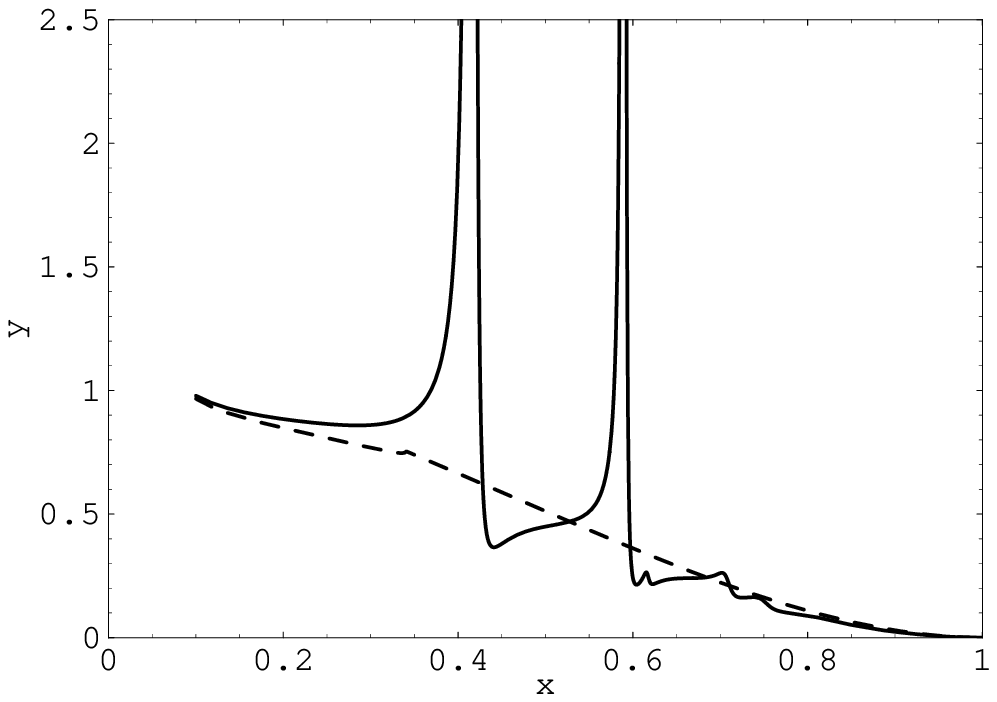}}
\caption{Spectra in inclusive modes: (left) Cut in the photon-energy
  spectrum in $\BtoXsgamma$.  (right) Differential $B\to
  X_s\ell^+\ell^-$ branching fraction as a function of $s=q^2/m^2_b\equiv
  m_{\ell^+\ell^-}^2/m_b^2$, including the effect of charm resonances in the
  Kr\"uger-Sehgal method (solid line). For comparison, the dashed curve shows the
  same quantity obtained within a purely partonic calculation at
  next-to-leading-log 
  precision, from Ref.~\cite{Beneke:2009az}.
\label{fig:bsllplot}
\label{toyspectrum}
}
\end{figure}

In the measurement of the inclusive mode $B \to X_s \gamma$ one
needs cuts in the photon-energy spectrum to suppress the background from
other $B$ decays (Fig.~\ref{toyspectrum}).

These shape-function effects were  taken into account in the experimental
analysis,  and the corresponding theoretical uncertainties due to this
model dependence are  reflected in the extrapolation error of  the
experimental results (see Section~\ref{sec:ex-xsgam}). The extrapolation is done 
from the experimental energy cut values down 
 to $1.6\;{\rm GeV}$ by use of three different theoretical
schemes~\cite{Benson:2004sg,Kagan:1998ym,Bosch:2004th,Buchmuller:2005zv}
for averaging.

Again constraining the analysis to the leading operator ${\cal O}_7$,
a cut around $1.6\,{\rm GeV}$ might not guarantee that a
theoretical description in terms of a local OPE is sufficient because of
the sensitivity to the scale $\Delta = m_b - 2
E_\gamma$~\cite{Neubert:2004dd}. A multiscale OPE with three
short-distance scales $m_b, \sqrt{m_b \Delta}$, and $\Delta$  has been 
proposed to connect the shape function and the local OPE region.
Recently, such additional perturbative cutoff-related effects have been 
calculated to NNLL precision by the use of SCET
methods~\cite{Becher:2006pu,Becher:2005pd,Becher:2006qw}.  Such
perturbative effects due to the additional scale are negligible for
$1.0\,{\rm GeV}$ but of order $3\%$ at $1.6\,{\rm
  GeV}$~\cite{Becher:2006pu}.  The size of these effects at $1.6\,{\rm
  GeV}$ is similar to the $3\%$ higher-order uncertainty in
the present NNLL prediction.  However, the numerical consistency of the
SCET analysis  has  recently been questioned~\cite{Misiak:2008ss}.  
Far away from the endpoint ($E_0=1.6$ GeV),
the logarithmic and nonlogarithmic terms cancel; the same result was
presented in Ref.~\cite{Andersen:2006hr}. Within the resummation of the
cutoff-enhanced logarithms this feature leads to an overestimate of the
$O(\alpha_s^3)$ terms~\cite{Misiak:2008ss}.  Further
work is needed to clarify this issue.

There is an alternative approach to the cut effects in the photon-energy
spectrum that is based on dressed gluon exponentiation and on the incorporation of  Sudakov
and renormalon resummations~\cite{Andersen:2005bj,Andersen:2006hr}.
The greater predictive power of this approach
is related in part to the assumption that nonperturbative power
corrections associated with the shape function follow the pattern of
ambiguities present in the perturbative calculation~\cite{Gardi:2006jc}.
In the future, these additional perturbative cut
effects could be analyzed and combined together with those already included
in the experimental average.

\paragraph{\boldmath$B \to X_s \ell^+ \ell^-$}

In the inclusive decay {\bf $B \to X_s \ell^+ \ell^-$}, 
the hadronic and dilepton invariant masses are independent
kinematical quantities.  A hadronic invariant-mass cut is imposed in the
experiments (see Section~\ref{sec:ex-xsll}).
The high-dilepton-mass region is not affected by this cut, but in the
low-dilepton mass region the kinematics with a jet-like $X_s$ and $m_X^2
\leq m_b \Lambda$ implies the relevance of the shape function.  A recent
SCET analysis shows that by using the universality of the shape
function, a  $10-30\%$ reduction in  the dilepton-mass spectrum can be
accurately computed. Nevertheless, the  effects of subleading shape functions
lead to an additional uncertainty of $5\%$~\cite{Lee:2005pk,Lee:2005pw}.  
A more recent analysis~\cite{Lee:2008xc}
estimates the uncertainties due to subleading shape functions more
conservatively. By scanning over a range of models of these functions,
one finds corrections in the rates relative to the leading-order result
 to be between $-10\%$ to $+10\%$ with equally large
uncertainties.  In the future it may be possible to
decrease such uncertainties significantly by constraining both the
leading and subleading shape functions using the combined $B \to
X_s\gamma$, $B \to X_u\ell\nubar$ and $B \to X_s \ell^+\ell^-$
data~\cite{Lee:2008xc}.

\subsection{Charmonium Resonance Contributions}
\label{charmoniumresonance}

One must  also consider the on-shell $c\cbar$ resonances,  which have to be taken
out.  Whereas  in the decay $B \rightarrow X_s \gamma$ the
intermediate $\psi$ background, namely ${B} \to \psi X_s$
followed by $\psi \to X' \gamma$, is suppressed for the high-energy cut
$E_\gamma$ and can be subtracted from the $B \rightarrow X_s
\gamma$ decay rate, the $c\cbar$ resonances show up as large peaks in
the dilepton-invariant mass spectrum in the decay $B \rightarrow
X_s \ell^+ \ell^-$.

As discussed in Section~\ref{perturbativecorrections}, these resonances
can be removed by making appropriate kinematic cuts in the invariant mass
spectrum.  However, nonperturbative contributions away from the
resonances within the perturbative windows are also important. In the
KS approach~\cite{Kruger:1996cv,Kruger:1996dt}
one absorbs
  factorizable long-distance charm rescattering effects (in which the
$B \to X_s c\cbar$ transition can be factorized into the product of
$\sbar b$ and $c\cbar$ color-singlet currents) into the matrix element
of the leading semileptonic operator ${\cal O}_9$.  Following the inclusion of 
nonperturbative corrections scaling with $1/m_c^2$, the KS approach
avoids double-counting.  For the integrated branching fractions one
finds an increase of $(1-2)\%$ in the low-$q^2$ region due to the
KS effect, whereas  in the high-$q^2$ region the increase  is well below the
uncertainty due to the $1/m_b$ corrections.  As shown in
Fig.~\ref{fig:bsllplot}, the integrated branching fraction is dominated
by this resonance background which exceeds the nonresonant charm-loop
contribution by two orders of magnitude.  
This feature should not be
misinterpreted as a striking failure of global parton-hadron duality~\cite{Beneke:2009az},
which postulates that the sum over the hadronic final states, including
resonances, should be well approximated by a quark-level
calculation~\cite{Poggio:1975af}.  Crucially, the
charm-resonance contributions to the decay $B \to X_s \ell^+\ell^-$
are expressed in terms of a phase-space integral over the absolute
square of a correlator.  For such a quantity global quark-hadron
duality is not expected to hold.  Nevertheless, 
local quark-hadron duality (which, of course, also implies global duality) 
  may be reestablished by resumming Coulomb-like
interactions~\cite{Beneke:2009az}.

\subsection{Soft Collinear Effective Theory for Exclusive Decays}
\label{softcollinear}

The Wilson coefficients of the weak effective Hamiltonian are
process-independent and therefore can be used directly in the
description of exclusive modes. It is computing of the hadronic
matrix elements between meson states that  is  difficult  in the
case of exclusive modes and that limits the theoretical precision.  The na\"{\i}ve
approach consists of  writing the amplitude $A \simeq C_i (\mu_b) \langle
{\cal O}_i (\mu_b) \rangle$ and parameterizing $\langle {\cal O}_i
(\mu_b) \rangle$ in terms of form factors. A substantial improvement can
be
obtained by using the QCDF method~\cite{Beneke:1999br,Beneke:2000ry,Beneke:2001ev} and its
field-theoretical formulation, the 
SCET method~\cite{Bauer:2000ew,Bauer:2000yr,Bauer:2001ct,Bauer:2001yt,Beneke:2002ph,Hill:2002vw}.
These methods form  the basis of the up-to-date predictions of exclusive
$B$ decays.  Within this framework one can show that, even if the form
factors were known with infinite precision, the description of exclusive
decays would be incomplete due to the existence of so-called
nonfactorizable strong interaction effects that do not correspond to
form factors.

The QCDF and SCET methods were first systematized for nonleptonic
decays in the heavy quark limit.  In contrast to the HQET,
SCET does not correspond to a local operator
expansion. Whereas  HQET is applicable to $B$ decays  if  the energy
transfer to light hadrons is small, for example to $B \rightarrow D$
transitions at small recoil to the $D$ meson, HQET is not applicable
 if  some of the outgoing, light particles have momenta of order $m_b$.
If so,  one faces a multi-scale problem that can be tackled within SCET.
In this case,  there are three relevant scales: (a) $\Lambda = {\rm few}
\times \Lambda_{\rm QCD}$, the soft scale set by the typical
energies and momenta of the light degrees of freedom in the
hadronic bound states; (b) $m_b$, the hard scale set by both  the
heavy $b$ quark mass and the energy of the final-state
hadron in the $B$ meson rest frame; and (c) the hard-collinear scale
$\mu_{\rm hc}=\sqrt{m_b \Lambda}$, which appears through interactions between
the soft and energetic modes in the initial and final states.  The dynamics
of hard and hard-collinear modes can be described perturbatively in the
heavy quark limit $m_b \to \infty$. Thus, SCET describes $B$ decays to
light hadrons with energies much larger than their masses, assuming that
their constituents have momenta collinear to the hadron momentum.

\paragraph{\boldmath$B\to K^*\gamma$ and $B\to\rho\gamma$}

The application of the QCDF
formalism to radiative and semileptonic decays was   first   
proposed in Ref.~\cite{Beneke:2000wa}.  For $B\to K^*\gamma$, or more generally for $B\to V\gamma$, where
$V$ is a light vector meson, the QCDF formula for the
hadronic matrix element of each operator of the effective Hamiltonian 
in the heavy quark limit and to all orders in $\alpha_s$ reads
\begin{equation}
\label{BVgamma}
\langle V \gamma |\, {\cal O}_i\, | B \rangle 
=  T^{I}_i\, F^{B\to V_\perp} 
+ \int_0^\infty{d\omega\over\omega}\,\phi_{B}(\omega)\, \int_0^1\,du\, 
\phi_{V_\perp}(u)\,T^{II}_i(\omega,u)\,. 
\end{equation}
This formula separates the process-independent nonperturbative quantities
$F^{B\to V_\perp}$, a form factor evaluated at maximum recoil ($q^2=0$),
and $\phi_{B}$ and $\phi_{V_\perp}$, the light-cone distribution amplitudes
(LCDAs) for the heavy and light mesons, respectively, from
the perturbatively calculable quantities $T^{I}$ and $T^{II}$.  The latter 
correspond to vertex and spectator corrections, respectively, and
have been calculated to
$O(\alpha_s^1)$~\cite{Beneke:2001at,Bosch:2001gv,Ali:2001ez,Descotes-Genon:2004hd}.
More recently, some $\alpha_s^2$ terms were also 
presented~\cite{Ali:2007sj}.

Light-cone wave functions of pseudo-scalar and vector mesons that  enter
the factorization formula have been studied in detail through the use of
light-cone QCD sum rules~\cite{Braun:1988qv,Braun:1989iv,Ball:1998sk,Ball:1998ff}. 
However, not much is known about the $B$ meson LCDA,
whose first negative moment enters the
factorized amplitude at $O(\alpha_s)$. Because this moment also enters the
factorized expression for the $B\to \gamma$ form factor, it
might be possible to extract its value from measurements of decays such as  
$B\to \gamma e \nu$, if the power corrections are under
control.

The QCDF formula also includes an important simplification
in the form factor description. The $B\to V$ form factors at large recoil
have been analyzed in
SCET~\cite{Bauer:2002aj,Beneke:2003pa,Lange:2003pk}
and are independent of the Dirac structure of the current
in the heavy quark limit; as a consequence, all $B\rightarrow
V_\perp$ form factors reduce to a   single form factor up to factorizable
corrections in the heavy quark and large energy
limits~\cite{Charles:1998dr,Beneke:2000wa}.

Field-theoretical methods 
such as  SCET make it possible  to reach a deeper
understanding of the QCDF approach.  The various momentum
regions are represented by different fields in the effective field
theory.  The hard-scattering kernels $T^{I}$ and $T^{II}$ can be shown
to be Wilson coefficients of effective field operators. Using SCET one
 can  prove the factorization formula to all orders in $\alpha_s$
and to  leading order in $\Lambda/m_b$~\cite{Becher:2005fg}.  QCD is
matched on SCET in a two-step procedure that separates the hard scale 
$\mu \sim m_b$ and then the hard-collinear scale $\mu\sim\sqrt{\Lambda m_b}$
from the hadronic scale $\Lambda$.  The vertex
correction term $T^I$ involves  the hard scales, whereas  the
spectator scattering term $T^{II}$ involves both the hard
and the hard-collinear scales.  This is why  large
logarithms have to be resummed~\cite{Becher:2005fg}, which can be done
most efficiently in SCET.

In principle, the field-theoretical framework of SCET allows one to go beyond
the leading-order result in $\Lambda/m_b$~\cite{Feldmann:2004mg}.
However, a breakdown of factorization is
expected at that   order~\cite{Beneke:2001ev}.  For example, in the
analysis of $B\to K^* \gamma$ decays at subleading order, an infrared
divergence is encountered in the matrix element of ${\cal
  O}_8$~\cite{Kagan:2001zk}.  In general, power corrections involve
convolutions,  which turn out to be divergent.  Currently, no solution to
this well-analyzed problem of end-point divergences within power
corrections is available~\cite{Becher:2003qh,Beneke:2003pa,Arnesen:2006vb}.  Thus, within
the QCDF/SCET approach, a general, quantitative method to estimate the
important $\Lambda/m_b$ corrections to the heavy quark limit is missing,
which significantly limits the precision in phenomenological  applications.

Nevertheless, some very specific power corrections are still 
computable and are often numerically important.  Indeed, this is the
case for the annihilation and weak exchange amplitudes in $B\to \rho
\gamma$.  The annihilation contributions also
represent the leading contribution to isospin
asymmetries~\cite{Kagan:2001zk}.  All these corrections are included
in recent analyses of radiative exclusive
decays~\cite{Ali:2004hn,Bosch:2004nd,Beneke:2004dp}.  Moreover, the
method of light-cone QCD sum rules can help provide estimates of
  such  unknown subleading terms. For example, power corrections for the
indirect $CP$ asymmetries in $B \to V \gamma$ decays have been analyzed in
this manner~\cite{Ball:2006eu}.

\paragraph{\boldmath$B\to K^{(*)}\ell^+ \ell^-$}

There is also a factorization formula for the exclusive semileptonic $B$
decays, such  as  $B \to K^* \ell^+ \ell^-$, that are  analogous to the one for the
radiative decay $B \to K^*
\gamma$~\cite{Beneke:2001at,Beneke:2004dp}. The simplification due to
form factor relations is even more drastic.  The hadronic form factors
can be expanded in the small ratios $\Lambda/m_b$ and $\Lambda/E$, where
$E$ is the energy of the light meson.  If we neglect  corrections of order
$1/m_b$ and $\alpha_s$, the seven a priori independent $B\to K^*$
form factors reduce to two universal form factors $\xi_{\bot}$ and
$\xi_{\|}$~\cite{Charles:1998dr,Beneke:2000wa}.  This reduction makes it possible
to design interesting ratios of observables in which any soft form
factor dependence cancels out for all dilepton masses $q^2$ at leading
order in $\alpha_s$ and $\Lambda/m_b$~\cite{Egede:2008uy}.

The theoretical simplifications of the QCDF/SCET approach are restricted
to the kinematic region in which the energy of the $K^*$ is of the order
of the heavy quark mass; that is, $q^2 \ll m_B^2$. Moreover, the influences
of very light resonances below $1\;{\rm GeV}^2$ question the QCDF
results in this  region. In addition, the longitudinal
amplitude in the QCDF/SCET approach generates a logarithmic divergence
in the limit $q^2 \rightarrow 0$, which indicates problems in the theoretical
description below $1\;{\rm GeV}^2$~\cite{Beneke:2001at}.  Thus, the
factorization formula applies well in the dilepton mass
range $1\; {\rm GeV}^2 < q^2 < 6\; {\rm GeV}^2 $.

Clearly, the  QCDF and SCET methods are also applicable to
 the phenomenologically  important semileptonic decays such as
$B \to K
\ell^+\ell^-$~\cite{Beneke:2001at,Bobeth:2007dw},
 $B \to \rho \ell^+\ell^- $ \cite{Beneke:2004dp}, and $B_s \to \phi \ell^+\ell^-$.
The decay mode into a pseudoscalar is  analogous to the decay mode  into
a longitudinal vector meson.

                   \section{EXPERIMENTAL TECHNIQUES}
                   \label{sec:ex-techniques}

The $\Upsilon(4S)$ resonance produced by the $\epem$ collision at the
$B$ factories provides a clean sample of $\BZ\BB$ and $\BP\BM$ meson
pairs as well as strong kinematical constraints that are otherwise unavailable,
particularly at hadron colliders.  The main background is from
continuum light quark pair production ($\epem\to\qqbar$, $q=u,d,s,c$),
which has a cross section only three times larger than that of $\BBbar$ production.
Radiative and electroweak penguin $B$ decays are efficiently measured
at the $B$ factories thanks to their clear signatures:
a high-energy photon and a lepton pair, respectively.

\subsection{Exclusive $B$ Reconstruction}

A $B$ meson decaying into an exclusive final state is reconstructed by
measuring all long-lived decay products ($\pi^\pm$, $K^\pm$, $e^\pm$,
$\mu^\pm$ and $\gamma$), selecting intermediate states of certain
invariant masses, and calculating two standard variables: the beam-energy constrained mass
$\Mbc=\sqrt{s/4-|\pB|^2}$ (also referred to as the beam-energy
substituted mass, $\Mes$) and the energy difference
$\DeltaE=\EB-\sqrt{s}/2$.  Here, $\sqrt{s}/2$ is the beam energy, and
$\pB$ and $\EB$ are the momentum and energy, respectively, of the reconstructed $B$
meson candidate in the $\Upsilon(4S)$ rest frame.

\begin{figure}[ht]
  \centerline{
    \myeps[0.45]{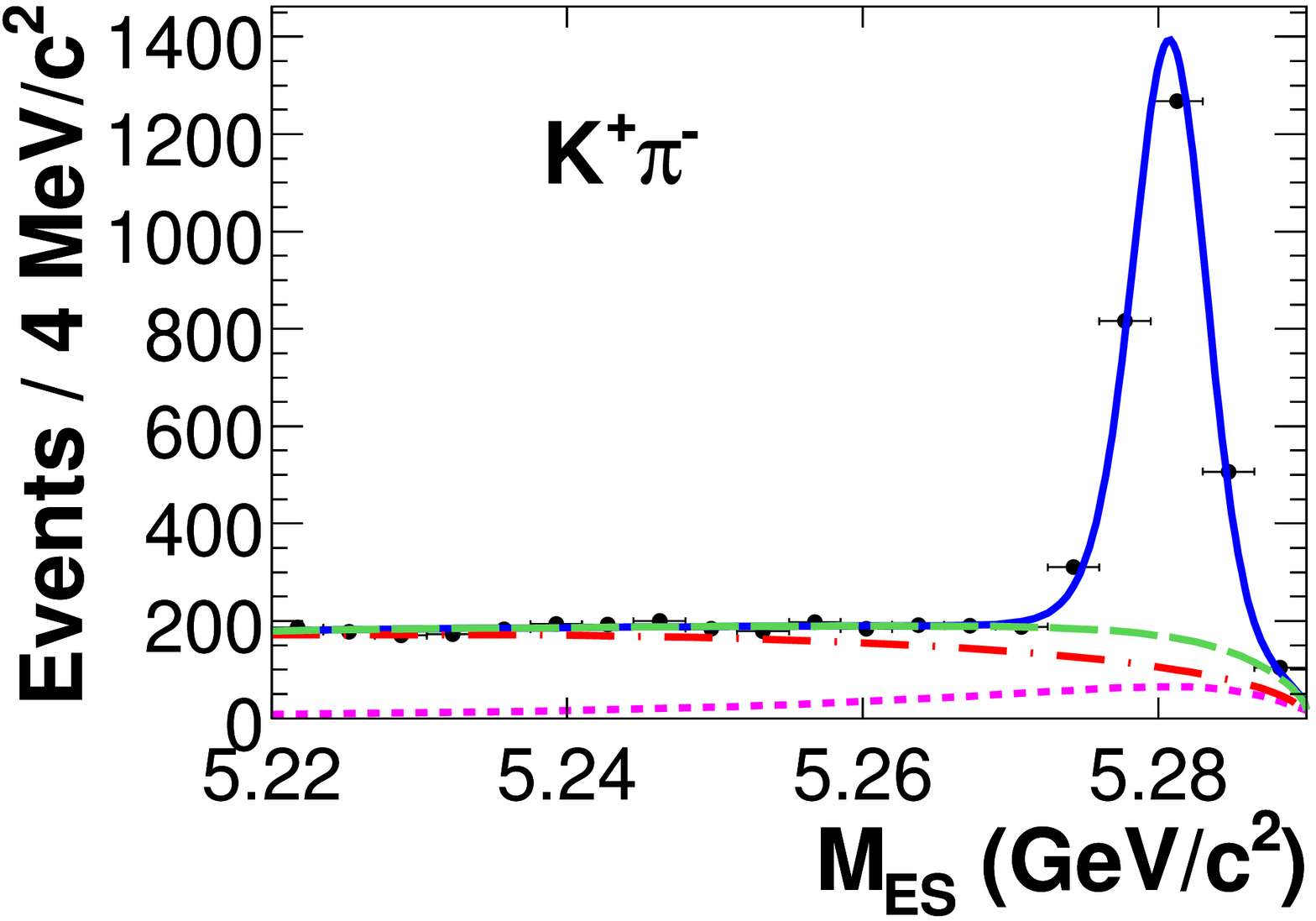}
    \myeps[0.45]{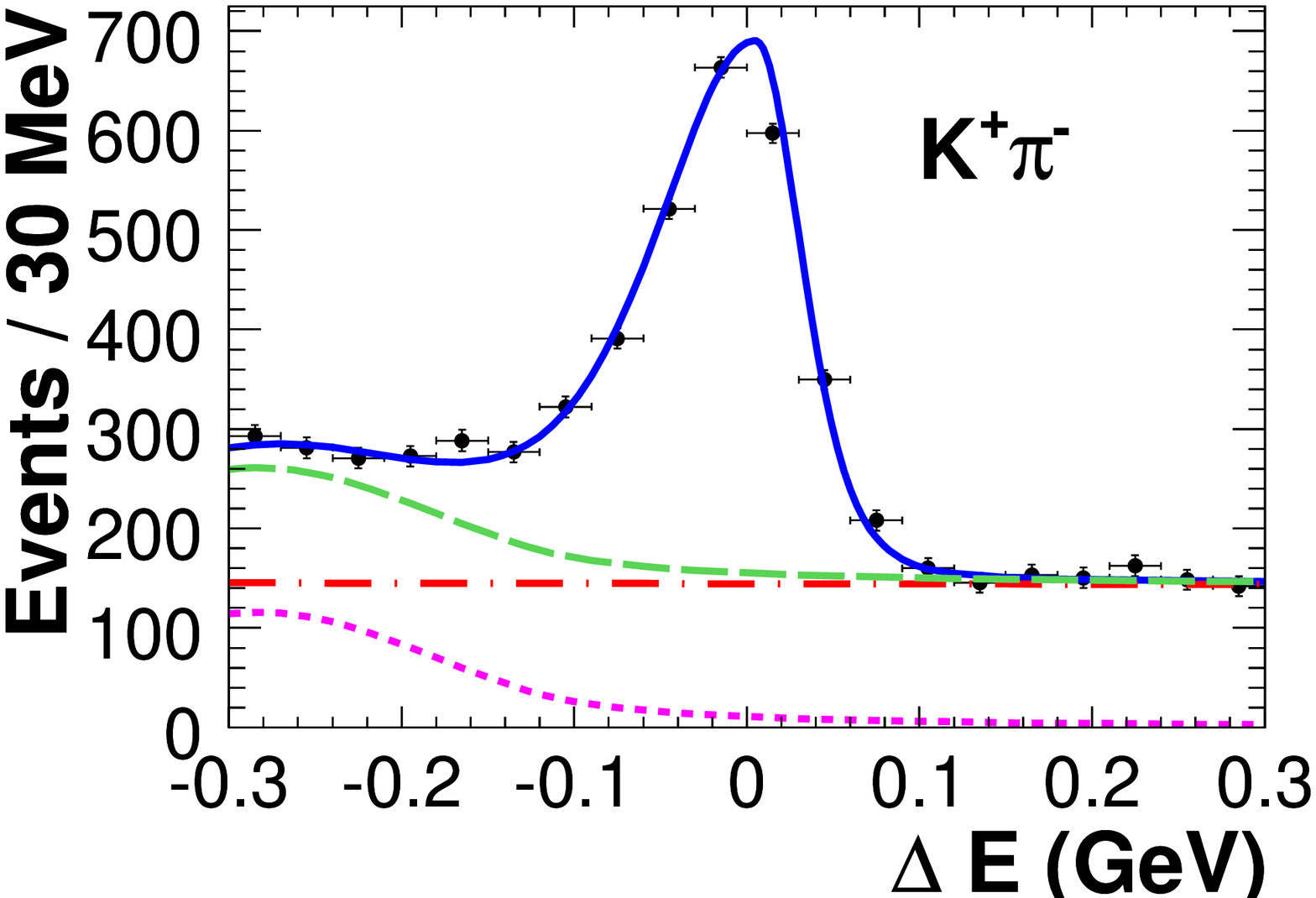}
  }
  \caption{Example of $\Mes$ ($\Mbc$) and $\Delta E$ for $\BtoKstarzerogamma$
    by BaBar (from Ref.~\cite{babar-kstgam}).}
  \label{fig:babar-kstgam}
\end{figure}

$\Mbc$ has a peak at the $B$
meson mass and $\DeltaE$ has a peak at zero (Fig.~\ref{fig:babar-kstgam}).  The resolution of $\Mbc$
is significantly better than that of $\DeltaE$, as the former is
dominated by the spread of the beam energy, whereas the latter is dominated by the
detector resolution.  The $\DeltaE$ variable is sensitive to
misreconstructed background events, whereas $\Mbc$ has little separation
power for them.  When a kaon is misidentified as a pion, $\DeltaE$
shifts by approximately 50 MeV, and when a low momentum pion is added or
missed, $\DeltaE$ shifts by more than the pion mass.  An exclusive
$\btodgamma$ final state is thus separated from a similar $\btosgamma$
state with $\DeltaE$, but this separation is marginal due to the photon energy
resolution of the electromagnetic calorimeter.
Therefore, pion to kaon
separation is crucial for the measurement of the suppressed
$\btodgamma$ processes.

Background events due to random combinations of particles are also
reduced by correctly identifying particle species.  For the $\btosll$
processes, electrons and muons are almost completely separated from the more
abundant hadrons.  In addition, various techniques based on the
event topology can be applied to suppress the background from continuum
$\qqbar$ events.

\subsection{Inclusive Measurement Techniques}

A fully inclusive measurement of $\BtoXsgamma$, in which the system
recoiling against the emitted photon is not reconstructed, has been
performed at the $B$ factories thanks to the clean environment. The dominant
background photon sources are (a) the copiously produced
$\piZ\to\gamma\gamma$ decays, (b) $\eta\to\gamma\gamma$ to a lesser extent,
and (c) other secondary and initial-state radiation photons in continuum
$\qqbar$ events.  These contributions can be safely subtracted because they are 
measured in events taken 60 or 40 MeV below the $\Upsilon(4S)$
resonance (i.e., off resonance).  Here, small corrections due to the
center-of-mass energy difference are applied to the production
cross section and the reconstruction efficiency.  
To avoid sacrificing  the $B$ decay sample for other studies,
the size of the
off-resonance data sample is only $\sim$10\% of the on-resonance sample
from both Belle and BaBar
and is the dominant source of
statistical and systematic errors (CLEO collected one-third of the
sample as off-resonance).  The second severe background source
arises from similar secondary photons from $B$ decays.  These contributions are
subtracted from the expected photon spectrum on the basis of measured $\piZ$
and $\eta$ spectra from $B$ decays, various control samples, or Monte Carlo
simulation.

An alternative technique is to measure as many exclusive modes as
possible and then calculate their sum (i.e., the sum-of-exclusive method).  Exclusive
branching fractions measured to date do not saturate the inclusive process,
but one can still infer the total branching fraction by estimating the
fraction of unmeasured modes of typically $\sim$45\% (or $\sim$30\% if
$\KL$ modes are accounted for by corresponding $\KS$ modes) using
simulated hadronization processes.  In the simulation, a light quark
pair is generated according to the SM mass spectrum and final-state
hadrons are produced by the PYTHIA program~\cite{pythia}.  This method also provides
direct information about the $B$ meson.  For example, the $B$ meson momentum
defines the $B$ meson rest frame, and charge and flavor information
allows $CP$- and isospin-asymmetry measurements.  So far, the sum-of-exclusive
method is the only way to perform inclusive measurements of $\BtoXsll$
and $\BtoXdgamma$ decays.

Another potentially definitive method is the so-called $B$-reco
technique, in which the other $B$ meson is fully reconstructed,
thereby allowing the target $B$ decay to be measured in a very clean
environment.  The efficiency is as low as a fraction of a percent, and
will be more important in future experiments.

               \section{PRESENT THEORETICAL PREDICTIONS}
               \label{sec:th-predictions}

Theoretical predictions have significantly improved in the past decade
along with the development of the theoretical tools. There have also been 
improvements in the relevant experimental input quantities,  as discussed
below.

\subsection{Inclusive Penguin Decays}
\label{inclusivepenguins}
The inclusive radiative and electroweak penguin modes offer  theoretically clean 
observables because nonperturbative corrections are small and well under control. 
This assessment also applies to  the branching fraction of $B \to X_d \gamma$ mode as discussed in 
Section~\ref{hadronicpower}. 

\paragraph{Inclusive \boldmath$\BtoXsgamma$}
The stringent bounds obtained from $\BtoXsgamma$ on various nonstandard
scenarios are a clear example of the importance of clean FCNC
observables for  discriminating NP models.

The branching fraction for $B\to X_q \gamma$ ($q=s,d$) can be
parameterized as
\begin{equation}
{\cal B} (B \to X_q \gamma)_{E_\gamma > E_0}
=  {\cal B}  (B \to X_c e \nubar)_{\rm exp} \, {6 \alpha_{\rm em} \over \pi C} \,
\left| V_{tq}^* V_{tb}^{}\over V_{cb}^{}\right|^2 
\, \Big[ P(E_0) + N(E_0) \Big] ,  \\
\label{br}
\end{equation}
where $\alpha_{\rm em} = \alpha_{\rm em}^{\rm on shell}$~\cite{Czarnecki:1998tn}, $C = |V_{ub}|^2 / |V_{cb}|^2 \,\times\, \Gamma[B\to X_c e
  \nubar] / \Gamma [ B\to X_u e \nubar] $ and $P(E_0)$ and
$N(E_0)$ denote the perturbative and  nonperturbative contributions,
respectively. The latter are  normalized to the charmless semileptonic
rate  to separate the charm  dependence.

 The first NNLL prediction, which is based on the perturbative calculations
discussed in Section~\ref{perturbativecorrections} and on the analyses
of nonperturbative corrections presented in Sections ~\ref{hadronicpower}
and \ref{nonperturbativecorrections}, for a photon-energy cut
$E_{\gamma} > 1.6\;{\rm GeV}$~\cite{Misiak:2006zs}, reads as: 
\begin{equation}\label{final1}\label{eq:xsgam-sm}
{\cal B}(B \to X_s \gamma)_{\rm NNLL} =  (3.15  \pm 0.23) \times 10^{-4}.
\end{equation}
The overall uncertainty consists of nonperturbative (5\%), parametric
(3\%), perturbative (scale) (3\%) and $m_c$-interpolation ambiguity
(3\%), which are  added in quadrature.
An additional scheme dependence in the
determination of the prefactor $C$ has been  found~\cite{Gambino:2008fj};  it is within the
perturbative uncertainty of $3\%$~\cite{Misiak:2008ss}.

Thus, the SM prediction and the experimental average (see
Section~\ref{sec:ex-xsgam}) are consistent at the $1.2 \sigma$ level.
This finding  implies very stringent constraints on NP models, such as (a) the 
bound on the charged Higgs mass in the two-Higgs doublet 
model~\cite{Ciuchini:1997xe,Borzumati:1998tg} ($M_{H^+} > 295{\rm GeV}$
at $95\%$ CL)~\cite{Misiak:2006zs} and (b)  the bound on the inverse
compactification radius of the minimal universal extra dimension model
($1/R > 600 {\rm GeV}$ at $95\%$ CL)~\cite{Haisch:2007vb}.  In
both cases, the bounds are much stronger than those  derived from other
measurements.  Constraints within various supersymmetric extensions have been 
analyzed in Refs.~\cite{Bertolini:1990if, Degrassi:2000qf,
  Carena:2000uj, Degrassi:2006eh, Borzumati:1999qt, Besmer:2001cj,
  Ciuchini:2002uv, Ciuchini:2003rg, Altmannshofer:2008vr} (for overviews
see~\cite{Hurth:2003vb,Altmannshofer:2009ne}).  Bounds on the little
Higgs model with $T$-parity have also been presented~\cite{Blanke:2009am}.
Finally, model-independent analyses in the effective field theory
approach without~\cite{Ali:2002jg} and with the assumption of minimal
flavor violation~\cite{ D'Ambrosio:2002ex, Hurth:2008jc} also show the
strong constraining power of the $B \to X_s \gamma$ branching
fraction.

\paragraph{Inclusive \boldmath$\BtoXdgamma$}

The theoretical predictions for the branching fraction ${\cal B}(B
\to X_d \gamma)$ for photon energies $E_\gamma>1.6$~GeV
read  as ~\cite{Hurth:2003dk,Hurth:2003pn}:
\begin{equation} 
\label{eq:xdgam-sm}
{\cal B} (B \to X_d \gamma)
 =  \Big( 1.38 \left. {}^{+0.14}_{-0.21}   \right|_{m_c \over m_b}
                     \pm 0.15_{\rm CKM}   \pm 0.09_{\rm param.} \pm 0.05_{\rm scale} \Big) \times 10^{-5} ,  
 \end{equation}
and
 \begin{equation}                   
\label{eq:BRratio}
{{\cal B} (B \to X_d \gamma) \over {\cal B} (B \to X_s \gamma)} =  \Big(3.82 \left. {}^{+0.11}_{-0.18}   \right|_{m_c \over m_b}
                     \pm 0.42_{\rm CKM}  \pm 0.08_{\rm param.} \pm 0.15_{\rm scale} \Big) \times 10^{-2}.
\end{equation}
These predictions are of NLL order. They are fully consistent with previous results~\cite{Ali:1998rr}.
A good part of the uncertainties
cancel out in the ratio.
The errors are dominated by CKM uncertainties,
and thus the measurement of ${\cal B}(B \to X_d \gamma)$ constrains the CKM
parameters.  This measurement  is also of specific interest with respect to
NP, because its CKM suppression by the factor
$|V_{td}|^2/|V_{ts}|^2$ in the SM may not hold in extended models.

\paragraph{Direct $CP$ Asymmetry}

Other important observables are the direct $CP$ asymmetries ($q=s,d$),
whose sign is always defined in terms of $b - \bbar$, or
\begin{equation}
\label{eq:acp-xqgam-def}
{\cal A}_{ CP} (\Bbar  \to X_q \gamma)   \equiv  {
    \Gamma(\Bbar \to X_q \gamma) - \Gamma(B \to X_{\qbar} \gamma) \over
    \Gamma(\Bbar \to X_q \gamma) + \Gamma(B \to X_{\qbar} \gamma)}.
\end{equation}
As first noted in Ref.~\cite{Kagan:1998bh}, the SM predictions are
almost independent from the photon energy cut-off and, for
$E_\gamma > 1.6$ GeV, read as ~\cite{Hurth:2003dk,Hurth:2003pn}
\begin{equation}
\label{eq:acp-xsgam-sm}
{\cal A}_{CP}(\Bbar \to X_s \gamma) =
\Big( 0.44 \left. {}^{+0.15}_{-0.10} \right|_{m_c \over m_b} 
   \pm 0.03_{\rm CKM}   \left. {}^{+0.19}_{-0.09} \right|_{\rm scale} \Big)\EM2,
\end{equation}
and
\begin{equation}   
{\cal A}_{CP}(\Bbar \to X_d \gamma) =
\Big( {-}10.2 \left. {}^{+2.4}_{-3.7} \right|_{m_c \over m_b}
  \pm 1.0_{\rm CKM} \left. {}^{+2.1}_{-4.4} \right|_{\rm scale} \Big)\EM2.
\end{equation}
The two $CP$ asymmetries are connected by the relative CKM factor
$\lambda^2 \, [(1-\rho)^2 + \eta^2]$.  The small SM prediction for the
$CP$ asymmetry in the decay $B \rightarrow X_s \gamma$ is a result of
three suppression factors: (a) $\alpha_s$    to have a strong phase;
(b) CKM suppression of order $\lambda^2$; and (c) GIM suppression of order
$(m_c/m_b)^2$, which reflects that in the limit $m_c = m_u$, any $CP$
asymmetry in the SM would vanish.

On the basis of CKM unitarity, one can derive the following $U$-spin relation
 between the un-normalized $CP$
 asymmetries~\cite{Soares:1991te}:
\begin{equation}
\left[\Gamma(\Bbar \to X_s \, \gamma)-\Gamma(B \to X_{\sbar } \, \gamma)\right] +
\left[\Gamma(\Bbar  \to X_d \, \gamma)-\Gamma(B \to X_{\dbar } \, \gamma)\right] =  0
\end{equation}
$U$-spin breaking effects can be estimated within the
HME (even beyond the partonic level), so  one arrives at
the following prediction for the total (or untagged) $\Bbar \to X_{s+d}
\gamma$ asymmetry~\cite{Hurth:2001yb,Hurth:2001ja}:
\begin{equation}
\label{eq:dcpv-s+d}
| \Delta {\cal B}(\Bbar  \to X_s \gamma) + \Delta {\cal B}(\Bbar  \to X_d \gamma) | \sim 1 \cdot 10^{-9} \; .
\end{equation}
Because this null test is based on the CKM unitarity, it represents
a clear test for new $CP$
phases beyond the CKM phase~\cite{Hurth:2001yb,Hurth:2001ja}.
 NP sensitivities of direct $CP$
asymmetries have been analyzed~\cite{Kagan:1998bh,Hurth:2003dk}.

\paragraph{Inclusive \boldmath$\BtoXsll$}

The decay $B \rightarrow X_s \ell^+\ell^-$ is particularly attractive
because it offers several kinematic observables.  The angular
decomposition of the decay rate provides three independent observables,
$H_T$, $H_A$ and $H_L$, from which one can extract the short-distance
electroweak Wilson coefficients that test for
NP~\cite{Lee:2006gs}:
\begin{equation}\label{eq:d3Gamma}
\frac{d^3\Gamma}{d q^2\,  d z}
= \frac{3}{8} \bigl[(1 + z^2) H_T(q^2)
+ 2(1 - z^2) H_L(q^2)
+  2 z H_A(q^2)
\bigr]
\,.\end{equation}
Here $z=\cos\theta_\ell$, $\theta_\ell$ is the angle between the negatively
charged lepton and the $\Bbar$ meson in the center-of-mass frame of
the dilepton system, and $q^2$ is  the dilepton mass squared.  $H_A$ is equivalent
to the forward-backward asymmetry, and the dilepton-mass spectrum is
given by $H_T + H_L$.  The observables depend on the Wilson coefficients
$C_7$, $C_9$ and $C_{10}$ in the SM. The present measurements of the
$B \to X_s \ell^+\ell^-$ already favor the SM-sign of the
coefficient $C_7$,  which is undetermined by the $B \to X_s \gamma$
mode~\cite{Gambino:2004mv}.

As discussed above, these observables are dominated by perturbative
contributions in the perturbative low- and high-$q^2$ windows which are
below ($1\;{\rm GeV}^2 < q^2 < 6\;{\rm GeV}^2$), and above ($q^2 > 14.4\;{\rm GeV}^2$) the $c \cbar$ resonances, respectively.  The present predictions are based
on the perturbative calculations to NNLL precision in QCD and to NLL
precision in QED (see Section~\ref{perturbativecorrections}).  
For the branching fraction in the low-$q^2$ region one arrives at~\cite{Huber:2005ig} 
\begin{equation}
\label{muonBR} 
{\cal B} (B\to X_s \ell^+\ell^-)_{\rm low} = \cases{
(  1.59  \pm 0.11 ) \times 10^{-6}  & $(\ell=\mu)$  \cr
(  1.64  \pm 0.11 ) \times 10^{-6}  & $(\ell=e) \, ,$    \cr}
\end{equation}
and for the high-$q^2$ region, one arrives at  ~\cite{Huber:2007vv}
\begin{equation}
\label{muonBRhighs} 
{\cal B} (B\to X_s \ell^+\ell^-)_{\rm high}  = \cases{
2.40 \times 10^{-7} \times (1^{+0.29}_{-0.26} ) & $(\ell=\mu)$ \cr
2.09 \times 10^{-7} \times (1^{+0.32}_{-0.30} ) & $(\ell=e) \, .$ \cr}
\end{equation}
As suggested in Ref.~\cite{Ligeti:2007sn}, normalizing the $B
\rightarrow X_s \ell^+ \ell^-$ decay rate in the high-$q^2$ region to
the semileptonic $B \rightarrow X_u \ell\nubar$ decay rate with
the same $q^2$ cut (Eq.~\ref{eq:zoltanR}), significantly reduces the
nonperturbative uncertainties~\cite{Huber:2007vv}:
\begin{equation} \label{muonR} 
{\cal R}(\hat q_0^2  =14.4{\rm GeV}^2)
 = \cases{
2.29 \times 10^{-3} \times ( 1 \pm 0.13) & $(\ell=\mu)$ \cr
1.94 \times 10^{-3} \times ( 1 \pm 0.16) & $(\ell=e) \, .$ \cr}
\end{equation}
The value of $q_0^2$ for which the forward-backward asymmetry vanishes,
\begin{equation}
\label{eq:muonzero}
(q_0^2)[X_s\ell^+\ell^-] = \cases{
( 3.50 \pm 0.12) \; {\rm GeV}^2  &  $(\ell=\mu)$ \cr
( 3.38 \pm 0.11) \; {\rm GeV}^2  &  $(\ell=e)\, ,$ \cr}
\end{equation}
is one of the most precise predictions in flavor physics and also
determines the relative sign and magnitude of the coefficients $C_7$ and
$C_9$~\cite{Huber:2007vv}.
However, unknown subleading nonperturbative corrections of order
$O(\alpha_s \Lambda/m_b)$, which are estimated to give an additional
uncertainty of order 5\%, have to be added in all observables of the $B
\to X_s \ell^+\ell^-$ mode (see Section~\ref{hadronicpower}).

In all predictions, it is assumed that there is no cut in the hadronic
mass region (see Section~\ref{nonperturbativecorrections}).
Furthermore, after including the NLL QED matrix elements, the electron
and muon channels receive different contributions due to terms involving
$\ln(m_b^2/m_\ell^2)$ (see Section~\ref{perturbativecorrections}). This
is the only source of the difference between these two channels. 
 All collinear photons are assumed to be included in the $X_s$
system, and the dilepton invariant mass does not contain any photons;
  in other words,  $q^2 = (p_{\ell^+} + p_{\ell^-})^2$.
Present experimental
settings at the $B$ factories are different,  and therefore  the theoretical
predictions have to be modified~\cite{Huber:2008ak}.

This difference in the settings also means that deviations from the SM prediction ($R_{X_s}^{\rm
  SM} =1$) in the muon-electron ratio
\begin{equation}
  \label{rk-def}
  R_{X_s} = {\Gamma(B\to X_s\mumu)_{[q_a^2,\,q_b^2]}
    / \Gamma(B \to X_s \epem)_{[q_a^2,\,q_b^2]}}
\end{equation}
can result from a different treatment of collinear photons in the
two modes.  This ratio is interesting because it is sensitive to the neutral
Higgs boson of two-Higgs-doublet models at large
$\tan\beta$~\cite{yan-rkkll,hiller-rkkll}, which  is also valid in
corresponding ratios $R_{K^{(*)}}$ of  exclusive modes: In the SM, one finds 
$R_K=1$, as well as  $R_{K^*}=0.75$ when integrated over all $q^2$,  including
$M_{\epem} < 2m_\mu$.

\subsection{Exclusive Penguin Decays}
\label{exclusivepenguins}
The exclusive penguin modes offer a larger variety of experimentally accessible observables than do 
the inclusive ones, but the nonperturbative uncertainties in the theoretical predictions 
are in general sizable.

\paragraph{\boldmath$\BtoKstargamma$ and $\Btorhogamma$}

The large hadronic uncertainties, which arise from the nonperturbative input of
the QCDF formula and from our limited knowledge of power
corrections,  do not allow precise predictions of the branching fractions
of exclusive modes. However,  within ratios of exclusive modes such as 
asymmetries,  parts of the uncertainties cancel out and one may hope for
higher precision.

The ratio $R_{\rm th} (\rho\gamma/K^* \gamma)$ [and similarly $R_{\rm
  th}(\omega\gamma/K^*\gamma)$] is given
by~\cite{Ali:2004hn,Bosch:2004nd,Beneke:2004dp,Ball:2006eu}.
\begin{equation}
R_{\rm th} ({\rho\gamma/ K^*\gamma})  = 
\frac{{\cal B}_{\rm th} (B \to \rho \gamma)}
     {{\cal B}_{\rm th} (B \to K^* \gamma)} =
S_\rho \left | \frac{V_{td}}{V_{ts}} \right |^2
\frac{(M_B^2 - m_\rho^2)^3}{(M_B^2 - m_{K^*}^2)^3} \,
\zeta^2 \, \left[ 1 + \Delta R (\rho/K^*) \right ],
\label{eq:Rth-rho/Ks} 
\end{equation}
where~$m_\rho$ is the mass of the $\rho$ meson;
$\zeta$~is the ratio of the transition form factors,
$\zeta=\Tbar_1^{\rho} (0)/\Tbar_1^{K^*} (0)$;
and $S_\rho = 1$ and~$1/2$ for the $\rho^\pm$ and $\rho^0$ mesons,
respectively.  The quantity $(1 +\Delta R)$ entails the explicit
$O(\alpha_s)$ corrections as well as the power-suppressed contributions.
These functions   also depend on CKM parameters, namely
$\phi_2\equiv\alpha=\arg(-\Vcb\Vcb^*/\Vtd\Vtb^*)$ and
$R_{ut}=|V_{ud}V_{ub}^*/V_{td}V_{tb}^*| $, and one finds
numerically~\cite{Beneke:2004dp} that
\begin{equation}
\Delta R(\rho^\pm/K^{* \pm}) =   \left\{ 1 - 2 R_{ut} \cos\phi_2 \,[0.24^{+0.18}_{-0.18}] 
   + R_{ut}^2 \, [0.07^{+0.12}_{-0.07}]  \right\},  \end{equation}
and
\begin{equation}
 \Delta  R(\rho^0/K^{*0}) = \left\{ 1 - 2 R_{ut}\cos\phi_2 \,[-0.06^{+0.06}_{-0.06}]
   + R_{ut}^2 \, [0.02^{+0.02}_{-0.01}] \right\}. 
\end{equation}
These results are consistent with the predictions given in Refs.~\cite{Ali:2004hn,Bosch:2004nd,Ball:2006eu}.
Obviously, the neutral mode is better suited for the determination of
$|V_{td}/V_{ts}|$ than is  the charged mode,  in which the function $\Delta R$
is dominated by the weak-annihilation contribution,  which leads to a
larger error.  The most recent determination of the ratio
$\zeta=\Tbar_1^{\rho} (0)/\Tbar_1^{K^*} (0)$ within the light-cone QCD
sum rule approach~\cite{Ball:2006nr}, $1/\zeta = 1.17 \pm 0.09$, leads
to the determination of $|V_{td}/V_{ts}|$ via Eq.~(\ref{eq:Rth-rho/Ks})
(see Section~\ref{sec:ex-rhogam}).  However, the experimental data on the
branching fractions of $B \to K^* \gamma$ and $B \to \rho \gamma$ calls
for a larger error on $\zeta$, if one assumes no large power
corrections beyond the known annihilation terms~\cite{Beneke:2004dp}.

\paragraph{Isospin Asymmetry in Radiative Decays}

Another  important observable is  the isospin breaking ratio  given by
\begin{equation}
  \label{eq:isospin-asym-def}
  \Delta_{0+}(\BtoKstargamma) = {
    \Gamma(\BtoKstarzerogamma) - \Gamma(\BtoKstarplusgamma) \over
    \Gamma(\BtoKstarzerogamma) + \Gamma(\BtoKstarplusgamma)},
\end{equation}
where the partial decay rates are $CP$-averaged.  In the SM
spectator-dependent effects enter only at the order $\Lambda/m_b$,  whereas
isospin-breaking in the form factors is  expected to be a negligible
effect. Therefore, the SM prediction is as small as
$O(5\%)$~\cite{Kagan:2001zk,Ali:2004hn,Bosch:2004nd,
  Beneke:2004dp,Ball:2006eu}.  The ratio is especially sensitive to NP
effects in the penguin sector, namely to the ratio of the two
effective couplings $C_6/C_7$.  The analogous isospin ratio in the
$\rho$ sector strongly depends on CKM parameters~\cite{Beneke:2004dp}:
\begin{equation}
  \Delta(\rho\gamma) =
  \frac{\Gamma(B^+ \to \rho^+\gamma)}{2\Gamma(
    B^0\to\rho^0\gamma)}-1 =  (-4.6 \, {}^{+5.4}_{-4.2} \big|_{\rm CKM} 
       {}^{+5.8}_{-5.6} \big|_{\rm had})\EM2.
\label{eq:isodef}
\end{equation}
The hadronic error is  due mainly to  the weak-annihilation
contribution,  to which a $50\%$ error is assigned.  

\paragraph{$CP$ asymmetries in Radiative Decays}

In the $CP$ asymmetries,  the uncertainties due to form factors cancel out to
a large extent. But both  the scale dependence and the dependence on
the charm quark of the next-to-leading-order predictions are rather large because the $CP$
asymmetries arise at  $O(\alpha_s)$ only.  Although  the direct $CP$
asymmetry in $B \to K^* \gamma$ is doubly Cabibbo suppressed and
expected to be very small within the QCDF/SCET approach, one finds $O(-10\%)$
predictions for the direct $CP$ asymmetries in the $B \to \rho \gamma$
mode~\cite{Bosch:2001gv,Ali:2004hn,Beneke:2004dp}. Because  the
weak-annihilation contribution does not contribute significantly here, the
neutral and charged modes are of  similar sizes~\cite{Beneke:2004dp}:
\begin{equation}
{\cal   A}_{CP}(\Bbar^0\to \rho^{0}\gamma) = 
   (-10.4 \, {}^{+1.6}_{-2.4} \big|_{\rm CKM} 
       {}^{+3.0}_{-3.6} \big|_{\rm had})\,\%
       \end{equation}
and
\begin{equation}
{\cal   A}_{CP}(B^-\to \rho^{-}\gamma) =    
   (-10.7 \, {}^{+1.5}_{-2.0} \big|_{\rm CKM} 
       {}^{+2.6}_{-3.7} \big|_{\rm had})\,\% . 
\label{directacp}
\end{equation}
Finally, we reiterate  that all predictions of exclusive observables within the QCDF/{\linebreak[1]}SCET approach may
receive further uncertainties due to the unknown power corrections. This possibility 
might be especially  important in the case of $CP$ asymmetries.

The time-dependent $CP$ asymmetry is given by two parameters, ${\cal
  S}_{CP}$ and ${\cal A}_{CP}$:
\begin{equation}
  \label{eq:tcpv-def}
  {\cal A}_{CP} (B\to f;\;\Delta t) = {\cal S}_{CP}  \sin(\Delta m\Delta t) + {\cal A}_{CP} \cos(\Delta
  m\Delta t),
\end{equation}
where ${\cal A}_{CP}$ represents the size of the direct $CP$ asymmetry
discussed above.\footnote{The symbol ${\cal C}_{CP}=-{\cal
    A}_{CP}$ is also often used.}  In hadronic decay modes such as $B\to J/\psi
\KS$, a large value of ${\cal S}_{CP}$
due to
the angle $\phi_1\equiv\beta=-\arg(\Vtd\Vtb^*/\Vud\Vub^*)$ of the
unitarity triangle has been established, 
and a similarly   large $CP$ asymmetry is expected for hadronic penguin
decays.  This asymmetry is suppressed in radiative penguin decays
because 
the photon helicities  are opposite between those from $B^0$ and
$\Bbar^0$ decays under the left-handed current of SM weak decays, and
they do not interfere in the limit of massless quarks.  This finding implies a
suppression factor of $m_s/m_b$ in the leading contribution to ${\cal
  S}_{CP}$     that is induced by the electromagnetic dipole operator ${\cal O}_7$:
\begin{equation}
{\cal S}^{\rm SM}_{CP} =- \sin 2\phi_1 \frac{m_s}{m_b} \left[2 +
  O(\alpha_s)\right] +{\cal S}^{{\rm SM},s\gamma g}  
\end{equation}
However, 
there are also additional contributions, ${\cal S}^{{\rm SM},s\gamma g}$ induced
by the process $b \to s \gamma g$ via operators other than ${\cal O}_7$~\cite{Grinstein:2004uu,Grinstein:2005nu}.
These corrections are not helicity-suppressed but are  power-suppressed.  A
\mbox{conservative} dimensional estimate of the contribution from  a
nonlocal SCET operator \mbox{series} leads to $|{\cal S}^{{\rm
    SM},s\gamma g}|
\approx 0.06$~\cite{Grinstein:2004uu,Grinstein:2005nu}, whereas  within a
QCD sum rule calculation, the contribution due to soft-gluon emission is
estimated to be ${\cal S}^{{\rm SM},s\gamma g} = - 0.005 \pm
0.01$~\cite{Ball:2006cva,Ball:2006eu} which leads to ${\cal S}^{\rm
  SM}_{CP} =-0.022\pm 0.015^{+0}_{-0.01}$.\footnote{This does not
  necessarily contradict a larger time-dependent $CP$ asymmetry of
  approximately
  $10\%$ within the inclusive mode found in
  Ref.~\cite{Grinstein:2004uu}, because the SCET estimate~\cite{Grinstein:2004uu,Grinstein:2005nu}  shows that the
  expansion parameter is $\Lambda/Q$.  Here $Q$ is the kinetic energy of
  the hadronic part. There is no contribution at leading order. Thus,
  the effect is expected to be larger for larger invariant hadronic
  mass. The $K^*$ mode must have the smallest effect, below the
  average $10\%$.}  The QCD sum rule estimates of power corrections,
namely long-distance contributions that arise from photon and soft-gluon emission
from quark loops~\cite{Ball:2006eu}, lead to analogous results for the
other radiative decay modes, such as $B \to \rho \gamma$~\cite{Ball:2006eu}.
If a large value of ${\cal S}_{CP}$ beyond the SM prediction is
observed, it  will  signal a new right-handed current
beyond the SM.

\paragraph{\boldmath$\BtoKstarll$}

The isospin asymmetry in the mode $B \to K^* \ell^+\ell^-$,    as in the radiative mode,                  is a
subleading $\Lambda/m_b$ effect, but the
dominant isospin-breaking effects can be calculated perturbatively, whereas   other $\Lambda/m_b$ corrections are simply  estimated. Thus, 
the exact uncertainty is difficult to estimate due to unknown power
corrections, but  the observable may still be useful in the NP search because of 
 its  high sensitivity  to specific Wilson
coefficients~\cite{Feldmann:2002iw}.

The decay $\Bbar^0 \to \Kbar^{*0} \ell^+ \ell^-$ (with $\Kbar^{*0} \to
K^-\pi^+$ on the mass shell)  is completely described by four independent
kinematic variables: the lepton-pair invariant mass squared, $q^2$, and
the three angles $\theta_\ell$, $\theta_{K}$, and $\phi$ (for their precise definitions, 
see Ref.~\cite{Egede:2008uy,Egedeneu}). 
Summing over the spins of
the final particles, the differential decay distribution can be written
as~\cite{Kruger:1999xa,Kruger:2005ep}
\begin{equation}
  \label{diff:four-fold}
  \frac{d^4\Gamma_{\Bbar}}{dq^2\,d\theta_\ell\, d
  \theta_K\, d\phi} = 
  \frac{9}{32 \pi} I(q^2, \theta_\ell, \theta_K, \phi) \sin\theta_\ell\sin\theta_K .
\end{equation}
By integrating two of the angles, one finds
  \begin{equation}
    \label{kstarll-thetak}
    \frac{d\Gamma^\prime}{d\theta_K} = 
    \frac{3\Gamma^\prime}{4} \sin\theta_K \left(
      2  F_L \cos^2 \theta_K + (1- F_L) \sin^2\theta_K
    \right),
  \end{equation}
and
  \begin{equation}
    \label{kstarll-thetal}
        \frac{d\Gamma^\prime}{d\theta_\ell} =  \Gamma^\prime \left(
      \frac{3}{4} F_L \sin^2\theta_\ell + 
      \frac{3}{8} (1- F_L) (1+\cos^2\theta_\ell) +
      A_{FB}  \cos\theta_\ell 
    \right)\sin\theta_\ell.
  \end{equation}
The observables appear linearly in the expressions so the fits can be
performed on data binned in $q^2$.  The fraction of longitudinal
polarization $F_L$ from the kaon angular distribution and the
forward-backward asymmetry $A_{FB}$ from the lepton angular distribution
are accessible this way.  The latter observable is defined as follows
 ($\theta_\ell$ is defined below Eq. \ref{eq:d3Gamma}):
\begin{equation}
 A_{FB}(q^2) \equiv 
 \frac{1}{d\Gamma/dq^2} \left(\,
\int_0^1 d(\cos\theta_\ell) \,\frac{d^2\Gamma}{dq^2 
d\cos\theta_\ell} - \int_{-1}^0 d(\cos\theta_\ell) \,\frac{d^2\Gamma}{dq^2 
d\cos\theta_\ell} \right) .
\end{equation}
The hadronic uncertainties of these two differential observables are
large.  However, the value of the dilepton invariant mass $q^2_0$, for
which the differential forward-backward asymmetry vanishes, can be
predicted in quite a clean way. In the QCDF approach at
leading order in $\Lambda/m_b$, the value of $q_0^2$ is free from
hadronic uncertainties at order $\alpha_s^0$. A dependence on the soft
form factor and on the light-cone wave functions of the $B$ and $K^*$
mesons appears  only at order $\alpha_s^1$.  At next-to-leading order one
finds~\cite{Beneke:2004dp}:
\begin{equation}
  q_0^2[K^{*0}\ell^+\ell^-] = 4.36^{+0.33}_{-0.31} \;\mbox{GeV}^2,\;\;\;\;
  q_0^2[K^{*+}\ell^+\ell^-] = 4.15^{+0.27}_{-0.27} \; \mbox{GeV}^2.
\end{equation}
The small difference is due to isospin-breaking power corrections.
However, an uncertainty due to unknown power corrections should be readded
to the theoretical error bars.  The zero is highly sensitive to
the ratio of the two Wilson coefficients $C_7$ and $C_9$. Thus, such a
measurement would have a huge phenomenological impact.

In the near future, a full angular
 analysis based on the four-fold differential decay rate in
Eq.~\ref{diff:four-fold} will become  possible. Such  rich information
would allow for the design of observables with specific NP sensitivity and
reduced hadronic uncertainties~\cite{Egede:2008uy,Egedeneu}.  These
observables would be
constructed in such a way that the soft form factor dependence would cancel
out at leading order for all dilepton masses, and they would
have much higher sensitivity to new right-handed
currents than would observables that   are already accessible via the
projection fits~\cite{Kruger:2005ep,Egede:2008uy,Egedeneu}.
In these optimized
observables, the unknown $\Lambda/m_b$ corrections would be the source of the
largest uncertainty.  Further detailed NP analyses of such  angular observables 
 have been   presented in Refs.~\cite{Bobeth:2008ij,Altmannshofer:2008dz}.  
A full angular analysis provides high sensitivity to various Wilson coefficients, but 
the sensitivity to new weak  phases is restricted~\cite{Egede:2009tp,Egedeneu}.

           \section{PRESENT EXPERIMENTAL RESULTS}
           \label{sec:ex-results}

The huge samples of $B$ meson decays collected by Belle and BaBar have made it
possible to fully explore the radiative penguin decays $\btosgamma$ and
$\btodgamma$, as well as the electroweak penguin decays $\btosll$.

\subsection{Inclusive $\BtoXsgamma$ Branching Fraction}
\label{sec:ex-xsgam}

\begin{figure}[ht]
  \centerline{
    \myeps[0.33]{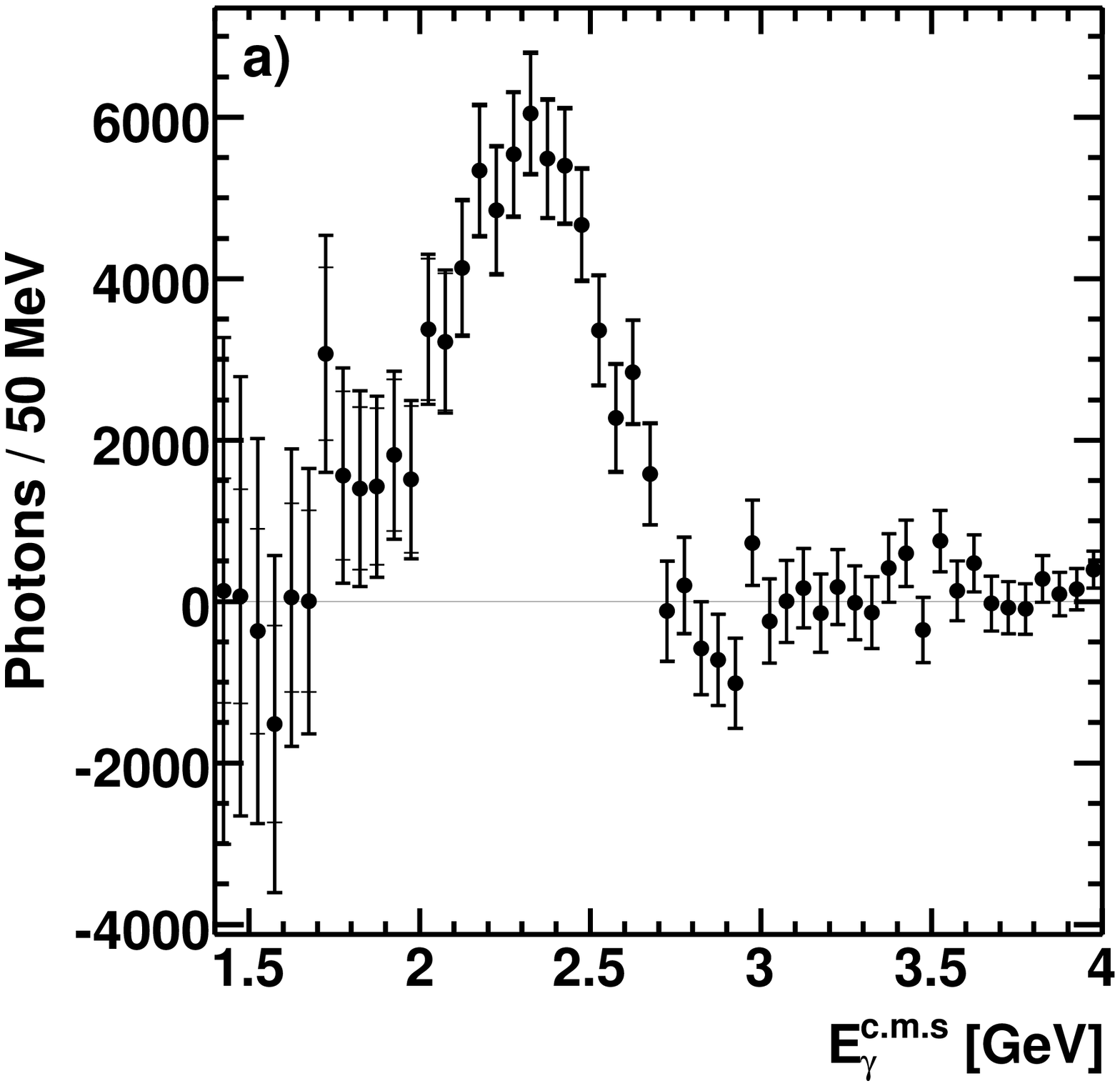}%
    \myeps[0.33]{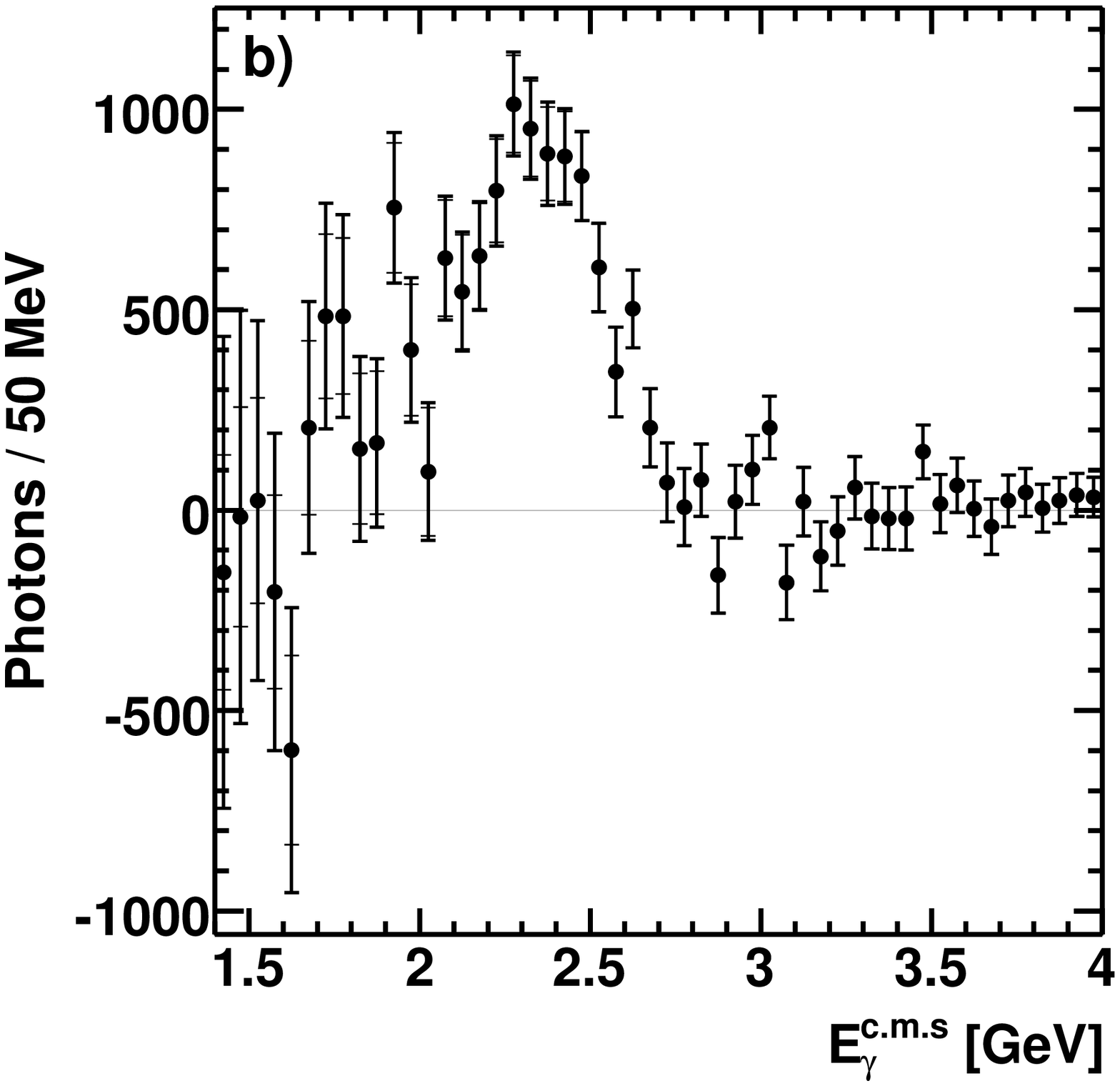}%
    \myeps[0.33]{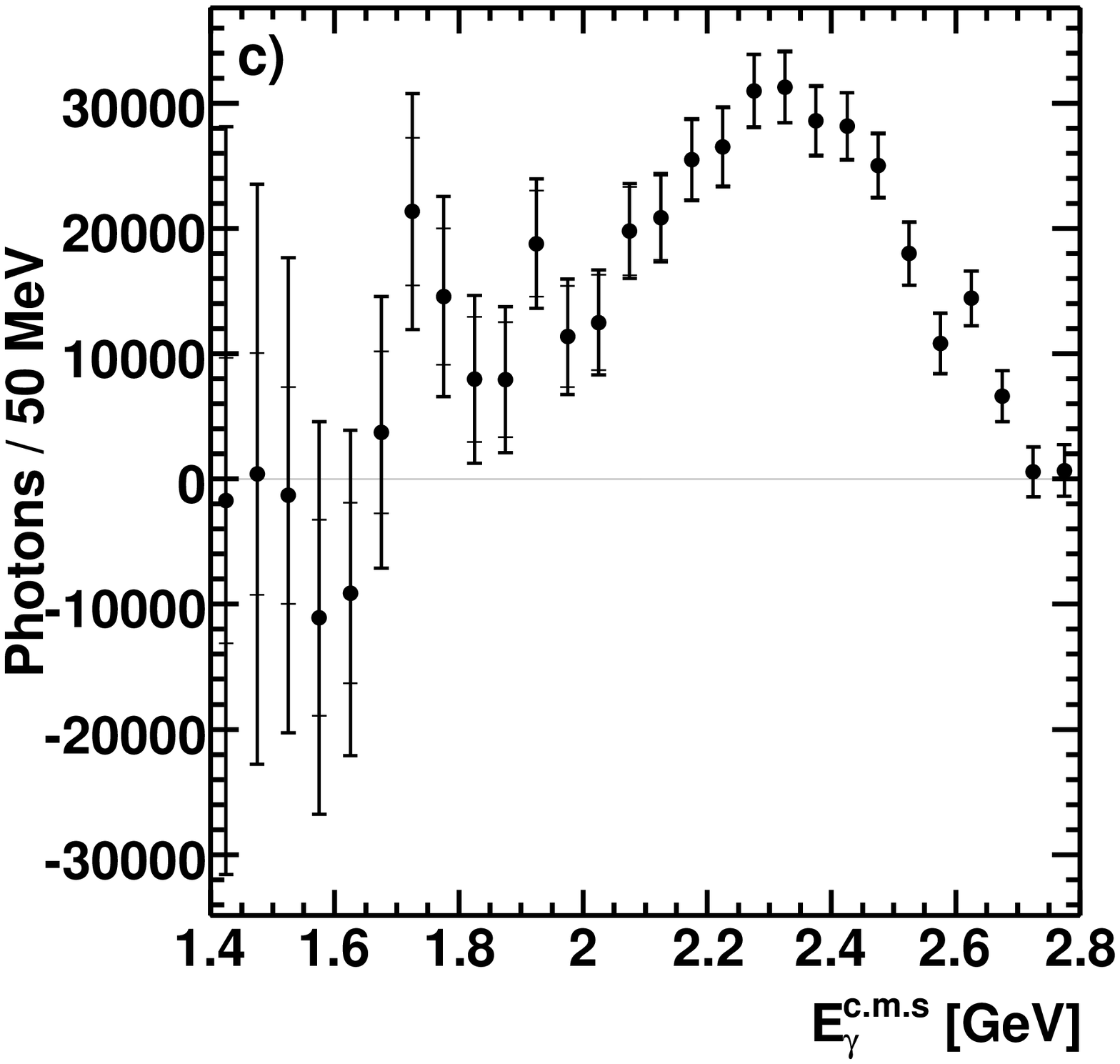}
    }
  \caption{Photon-energy spectrum for $\BtoXsgamma$, as measured by Belle
    (a) without lepton tag, (b) with a lepton tag, and (c) their average
    from Ref.~\cite{belle-xsgami}.}
  \label{fig:belle-xsgam}
\end{figure}

An experimental challenge is how to lower the minimum photon
energy to $1.6\GeV$ (see Section~\ref{nonperturbativecorrections}).
Before the construction of the $B$ factories, 
minimum photon energy of 2.0 GeV was required in the measurement by
CLEO~\cite{cleo-xsgami}.
BaBar has a minimum photon-energy requirement of 1.9 GeV based on 89
million $\BBbar$ pairs~\cite{babar-xsgami}, whereas Belle first reported
the result of 1.8 GeV with 152 million $\BBbar$~\cite{belle-xsgamo}.
Belle recently lowered the limit to 1.7 GeV by using 657 million $\BBbar$
pairs~\cite{belle-xsgami} (Fig.~\ref{fig:belle-xsgam}).  Belle measured
the branching fraction to be $\Br(\BtoXsgamma)=\BrBtoXsgamIBelle$ for
$\Egamma>1.7\GeV$, whereas BaBar measured it to be
$\BrBtoXsgamIBaBar$ for $\Egamma>1.9\GeV$.

The sum-of-exclusive technique and the
$B$-reco technique have been used by BaBar with a minimum 
photon-energy requirement of 1.9 GeV~\cite{babar-xsgamx, babar-xsgamb},
which
corresponds to a maximum recoil mass requirement of 2.8 GeV.  Belle
also made a sum-of-exclusive measurement using a very early
data set~\cite{belle-xsgamx}.

To calculate the average branching fraction based on the same
phase space and to compare it with theory predictions, the Heavy Flavor
Averaging Group (HFAG)~\cite{hfag} has made an extrapolation of the
branching fraction to the same minimum photon energy of
$1.6\GeV$~\cite{Buchmuller:2005zv}.  The extrapolation factor is
$0.985\pm0.004$ ($0.894\pm0.016$) for 1.7 (2.0) GeV.  The average
thereby obtained is
\begin{equation}
  \Br(\BtoXsgamma) = \BrBtoXsgamHFAG,
\end{equation}
where the first error is statistical and systematic combined, and the second
is due to the extrapolation.  The result is in agreement with the SM
prediction given in Eq.~\ref{eq:xsgam-sm}, and it provides stringent constraints on
NP, as discussed in Section~\ref{inclusivepenguins}.

\subsection{Exclusive Measurements of $\btosgamma$ Processes}

The recoil system of $\BtoXsgamma$ below 1.1 GeV is dominated by the
$K^*$ resonance, as the spin-0 state is forbidden.  Above 1.1 GeV, $X_s$
is a mixture of various resonant and nonresonant states and
therefore can usually be modeled as a continuum spectrum in the inclusive
$\BtoXsgamma$ analysis.  The $K^*$ signal makes it possible  to use
the channel for various studies, as discussed below.  The $B\to K^*\gamma$
branching fractions have been measured precisely by
Belle~\cite{belle-kstgam} and BaBar~\cite{babar-kstgam} and have been averaged by
HFAG to be
\begin{equation}
  \begin{array}{l}
  \Br(\BtoKstarzerogamma) = \BrBtoKstZgamHFAG \\
  \Br(\BtoKstarplusgamma) = \BrBtoKstPgamHFAG , \\
  \end{array}
\end{equation}
which corresponds to approximately 12\% of the total $\BtoXsgamma$ branching
fraction.  The SM predictions for the branching fraction have been
calculated by many groups. They are  consistent with the measured values
but have very large errors of 30\% to 50\%, which arise mainly from the uncertainty
of the $B\to K^*$ form factor~\cite{Beneke:2001at,Bosch:2001gv}. 

The resonant structure of the high mass $X_s$ system has also been explored.
So far, $B\to K_2^*(1430)\gamma$~\cite{belle-kxgam,babar-k2stgam}
and $B\to K_1(1270)\gamma$~\cite{belle-k1gam} have been measured, but
other decay channels such as $B\to K_1(1400)\gamma$ seem to have small
branching fractions and have not yet been observed.  In addition, many
multi-body final states have been measured, such as $B\to
K\pi\pi\gamma$~\cite{babar-kpipigam, belle-kxgam} (including $B\to
K^*\pi\gamma$ and $B\to K\rho\gamma$), $B\to
K\phi\gamma$~\cite{belle-kphigam,babar-kphigam}, $B\to
K\eta\gamma$~\cite{belle-ketagam,babar-ketagam}, $B\to
K\eta'\gamma$~\cite{belle-ketapgam}, and $B\to \Lambda
\pbar\gamma$~\cite{belle-pLgam}.

\subsection{$CP$ and Isospin Asymmetries in $\btosgamma$ Processes}

The measurement of the 
inclusive direct $CP$ asymmetry (Eq.~\ref{eq:acp-xqgam-def})
was performed by use of the
sum-of-exclusive method to tag the flavor of the $B$ candidate.  For
$B^0$ ($\Bbar^0$), only the self-tagging modes with a $K^+$ ($K^-$)
were used.  The measured asymmetry was corrected for a small dilution due
to the doubly misidentified pair of
a charged kaon and a pion.  The results, based on 152 and 383 million
$\BBbar$ samples by Belle and BaBar, are
$\AcpBtoXsgamXBelle$~\cite{belle-acpxsgam} and
$\AcpBtoXsgamXBaBar$~\cite{babar-acpxsgam}, respectively, and have been
averaged by HFAG to be
\begin{equation}
  \Acp(\BbartoXsgamma) = \AcpBtoXsgamHFAG.
\end{equation}
This is consistent with null asymmetry, and the size of the error is
still much larger than the SM precision (Eq.~\ref{eq:acp-xsgam-sm}).
Assuming the systematic error can be reduced along with the statistical
error, a data set two orders of magnitude larger would be more sensitive
to NP, although still insufficient to
measure the small $\Acp$ predicted by the SM.

In the exclusive $\BtoKstargamma$ channel, the direct $CP$ asymmetry is
also small but has less-understood theoretical uncertainties.
Experimentally it can be measured more precisely. The current HFAG average
is
\begin{equation}
  \Acp(\BtoKstargamma) = \AcpBtoKstgamHFAG,
\end{equation}
which is also consistent with null asymmetry.

In the fully inclusive measurement, flavor information is not
available for the signal side, but it can be obtained from the charge of the
lepton in the event if the other $B$ decays into a semileptonic final
state.
In this case, it is not possible to discriminate $\BtoXdgamma$ from
$\BtoXsgamma$ and the measured asymmetry corresponds to a combined one.
This combined asymmetry (note  the different normalization 
in comparison with Eq.~\ref{eq:dcpv-s+d}) has been  measured by BaBar to be
\begin{equation}
  \Acp(\BbartoXsdgamma) = \AcpBtoXsdgamIBaBar,
\end{equation}
which is consistent with null asymmetry but has a
much larger error than do the other two asymmetry measurements.

The measurement of the isospin asymmetry (Eq.~\ref{eq:isospin-asym-def}) is
another way to utilize the reconstructed $\BtoKstargamma$ events.
Here, the measured branching fractions are corrected by  the lifetime
ratio: $\tau_{B^+}/\tau_{B^0}$ $= 1.071\pm0.009$~\cite{pdg}.
Usually $B$ decay branching fractions are
quoted based on the assumption of $\Br(\Upsilon(4S)\to
B^+B^-)=\Br(\Upsilon(4S)\to B^0\Bbar^0)=0.5$, but the isospin-asymmetry
measurement is already precise enough to be affected by the difference
in these branching fractions.  Belle measures
$\Delta_{0+}(\BtoKstargamma)=\AIBtoKstgamBelle$ without this correction,
whereas BaBar measures $\Delta_{0+}(\BtoKstargamma)=\AIBtoKstgamBaBar$
using $\Br(\Upsilon(4S)\to B^+B^-)=0.516\pm0.006$ and
$\Br(\Upsilon(4S)\to B^0\Bbar^0)=0.484\pm0.006$~\cite{pdg}.  After
scaling the Belle result and including the CLEO result using the
aforementioned lifetime and production ratios, the na\"{\i}ve world average
is 
\begin{equation}
  \Delta_{0+}(\BtoKstargamma) = 0.062\pm0.027\,.
\end{equation}
This average is in agreement with the SM expectation.

%
%
%

A similar isospin asymmetry can be also measured for the inclusive
$\BtoXsgamma$ decay by use of the sum-of-exclusive method.  The result by BaBar
is
\begin{equation}
  \Delta_{0+}(\BtoXsgamma) = \AIBtoXsgamXBaBar,
\end{equation}
which is consistent with null asymmetry but is not yet as precise as
that for $\BtoKstargamma$.

The measurement of the time-dependent $CP$ asymmetry (Eq.~\ref{eq:tcpv-def}) for
$\btosgamma$ faces two experimental challenges.
First, the modes and statistics
that can be used for time-dependent $CP$ asymmetry measurements are
rather limited.  Although the $\BtoKstarzerogamma$ branching fraction is
not very small, only 1/9 of the events that decay into the
$\KS(\to\piP\piM)\piZ\gamma$ final state can be used.  Second, the $B$
meson decay vertex position has to be extrapolated from the displaced
$\KS\to\piP\piM$ vertex and the $\KS$ momentum vector. Therefore, 
the $\KS$ decays inside the vertex detector volume (55\%
in Belle, 68\% in BaBar) and the resulting vertex resolution is somewhat
degraded.

Because any $B\to P^0 Q^0\gamma$ final states [where $P^0$ and $Q^0$ are $CP$
eigenstates~\cite{atwood-tcpvpqgam}]  can be used, the $B\to\KS\piZ\gamma$
events with $M_{\KS\piZ}$ up to $1.8\GeV$ including $K_2^*(1430)$ were
measured by both Belle~\cite{belle-tcpvkspizgam}
and BaBar~\cite{babar-tcpvkspizgam} and have been averaged by HFAG as
\begin{equation}
  \begin{array}{l}
    \Scp(B\to\KS\piZ\gamma)=\ScpBtoKSpiZgamHFAG, \\
  \end{array}
\end{equation}
in  which the $\BtoKstarzerogamma$ contribution gives
$\Scp(\BtoKstargamma)=\ScpBtoKstZgamHFAG$.  

As additional channels, BaBar has measured
$\Scp(B\to\KS\eta\gamma)=\ScpBtoKZetagamBaBar$~\cite{babar-ketagam}, and Belle has measured
$\Scp(B\to\KS\rhoZ\gamma)=\ScpBtoKSrhoZgamBelle$~\cite{belle-tcpvksrhogam}.  The latter is
slightly diluted by the $B\to\Kstar^+\pi^-\gamma$ events, by a
factor which was measured to be $0.83\PM{0.19}{0.03}$, but is free
from the restriction of $\KS$ vertexing and has a 
statistical error comparable in size to that of the $B\to\KS\piZ\gamma$ mode.
Currently all results are compatible with null asymmetry with errors
that are still not small enough to provide nontrivial constraints on
right-handed currents, but this  observable  will be one of the best ways
to search for NP in future experiments.

\subsection{Measurements of $\btodgamma$ Processes}
\label{sec:ex-rhogam}

Three exclusive $\btodgamma$ decay modes are considered to be the
easiest modes to study the $\btodgamma$ process: $\Btorhoplusgamma$,
$\Btorhozerogamma$ and $\BZtoomegagamma$.  Although these modes
have been searched for since the beginning of Belle and BaBar, 
only in the later stage of these experiments were measurements of the
$\Btorhogamma$ modes established.  This is partly because of the large
$\BtoKstargamma$ background and partly due to  the huge continuum
background, which is more severe for modes without a kaon in the
final state.  
Therefore, large statistics and good particle identification
are essential; Belle has the advantage in the former, whereas BaBar leads in
the latter.

To gain statistics, these three modes have been combined by
assuming their na\"{\i}ve quark contents, through the use of $\Gamma(\Btorhoplusgamma)=2
\Gamma(\Btorhozerogamma)=2\Gamma(\BZtoomegagamma)$.  In the
latest measurements by both Belle~\cite{belle-rhogam} and
BaBar~\cite{babar-rhogam}, the $\Btorhozerogamma$ mode was measured
with more than $5\sigma$ significance and $\Btorhoplusgamma$ with more than $3\sigma$
significance,
whereas $\BZtoomegagamma$ remains unestablished with significance less
than $3\sigma$.  Using the
symbol $\BtoROgamma$ for the combined results that are adjusted for the
$\Btorhoplusgamma$ mode, the averaged branching fraction by HFAG becomes
\begin{equation}
  \begin{array}{l}
    \Br(\BtoROgamma) = \BrBtoROgamHFAG. \\
  \end{array}
\end{equation}
The results are consistent with the SM predictions.  However, these predictions are affected by
form factor uncertainties and do not have effective prediction power
for NP.

A more effective way to use these results is
to combine them with the $\BtoKstargamma$ measurements to determine
$|\Vtd/\Vts|$.
Using Eq.~\ref{eq:Rth-rho/Ks}, Belle and BaBar reported
the value of $|\Vtd/\Vts|$ to be $\absVtdoVtsBelle$ and
$\absVtdoVtsBaBar$, respectively, where the errors are experimental and
theoretical.  These results from the penguin diagrams
are in agreement with the determination from the box diagrams using
the ratio of the $B^0$ and $B_s^0$ mixing parameters $\Delta m_d/\Delta
m_s$, where $\Delta m_d$ is measured at the $B$ factories and $\Delta m_s$ at
the Tevatron.  The results are also in agreement with the more indirect determination by 
the unitarity triangle fit from other observables.  
This is a nontrivial test of the CKM scheme.
However, although the experimental errors are still larger than the
theoretical errors, the size of the theoretical error is unlikely to be
reduced.

A possible way to improve this situation utilizes the inclusive $B\to
X_d\gamma$ measurement with the sum-of-exclusive method.
BaBar has reconstructed the $X_d\gamma$ system in seven final
states
($\piP\piM\gamma$,
$\piP\piZ\gamma$,
$\piP\piM\piP\gamma$,
$\piP\piM\piZ\gamma$,
$\piP\piM\piP\piM\gamma$,
$\piP\piM\piP\piZ\gamma$,
$\piP\eta\gamma$)~\cite{babar-xdgam}
in the mass range $0.6 < M_{X_d} < 1.8\GeV$,  which covers approximately 50\% of
the total branching fraction.  To
reduce the uncertainty due to missing modes and phase space, $X_s\gamma$
modes were also measured in the corresponding seven final states
in the same mass range, where the first $\piP$ was replaced with $\KP$.
The ratio of the two inclusive branching fractions is
\begin{equation}
  {\Br(\BtoXdgamma)\over\Br(\BtoXsgamma)}=\rBtoXdoXsgamBaBar,
\end{equation}
which is converted to $|\Vtd/\Vts|=\absVtdoVtsBtoXdgamBaBar$, where the
theory error does not include the effect due to the limited mass range.
This result is also in good agreement with other determinations.

The direct $CP$ asymmetry for $\Btorhogamma$ can be as large as $\sim -10\%$ in
the SM, whereas the time-dependent $CP$ asymmetry is doubly suppressed due to
the photon helicity and the cancellation of the CKM element $\Vtd$. The latter 
appears in the mixing and in the $\btod$ penguin decay.
However, the $\rhoZ\to\piP\piM$ decay provides clear vertex information
for $\Btorhozerogamma$.  Both $CP$ asymmetries have been measured by
Belle~\cite{belle-rhogam,belle-tcpvrhogam} as
\begin{equation}
  \begin{array}{l}
  \Acp(\Btorhoplusgamma) = \AcpBtorhoPgamBelle, \\
  \Acp(\Btorhozerogamma) = \AcpBtorhoZgamBelle, \mbox{~and} \\
  {\Scp}(\Btorhozerogamma) = \ScpBtorhoZgamBelle. \\
  \end{array}
\end{equation}
So far the results are consistent with null asymmetry.  A nonzero
direct $CP$ violation may be measured earlier in $\Btorhogamma$ than in
$\BtoKstargamma$.

The isospin asymmetry (Eq.~\ref{eq:isodef}) is also expected to be as large as $\sim -5\%$ in $\Btorhogamma$.  
Belle measures $\Delta(\rho\gamma)=\DeltaBtorhogamBelle$~\cite{belle-rhogam},
and BaBar measures
$\Delta(\rho\gamma)=\DeltaBtorhogamBaBar$~\cite{babar-rhogam}; both measurements show
a large isospin asymmetry. The average by HFAG is
\begin{equation}
  \Delta(\rho\gamma) = \DeltaBtorhogamHFAG.
\end{equation}
A significant nonzero isospin asymmetry could indicate NP.

\subsection{Exclusive $\BtoKorKstarll$ Branching Fraction}

Despite their small branching fractions, the exclusive decay channels
$\BtoKorKstarll$ have been measured efficiently with small background at
Belle and BaBar, given that their final states are the same as those of $B\to
\Jpsi K^{(*)}$ for which the $B$ factories were designed.  Here, $K^{(*)}$ is
one of $K^+$, $\KS$, $K^{*+}$ and $K^{*0}$, and $\elel$ is either
$\epem$ or $\mumu$.

Electrons are identified by their energy deposit through an
electromagnetic shower in the calorimeter. The minimum momentum is
required to be greater than 0.4 GeV by Belle or 0.5 GeV by BaBar.  The
momentum of the bremsstrahlung photons that may be emitted by the
electrons are added to their momenta if they are found near the electron
direction.  Muons have to reach and penetrate into the outer muon
detectors and the minimum momentum is required to be 0.7 GeV by Belle or
1.0 GeV by BaBar.  The dilepton mass regions around $\Jpsi$ and
$\psi(2S)$ are vetoed.

The branching fractions, averaged over the lepton and kaon flavors and
integrated over the dilepton masses, assuming the SM distribution over
the vetoed $J/\psi$ and $\psi(2S)$, were measured by
Belle~\cite{belle-kstll} and BaBar~\cite{babar-kstll} and have been
averaged by HFAG as
\begin{equation}
  \begin{array}{l}
    \Br(\BtoKll)     = \BrBtoKllHFAG,\\
    \Br(\BtoKstarll) = \BrBtoKstllHFAG.
  \end{array}
\end{equation}
The results are consistent with SM expectations.
At present, the irreducible form factor uncertainty in the SM
calculations prevents these results from placing meaningful constraints
on NP.

A small subset of these combinations, $B^+\to K^+\mumu$ and
$B^0\to\KstarZ\mumu$, can be efficiently measured at hadron colliders. CDF has
reported the most precise measurements of these modes~\cite{cdf-kstll}.

\subsection{$\BtoKorKstarll$ Asymmetries and Angular Distributions}

The direct $CP$ and isospin asymmetries in $\BtoKorKstarll$ are also useful in the
search for NP.  The direct $CP$ asymmetries are
consistent with null values, $\Acp(B\to K^+\elel)=\AcbBtoKPllHFAG$
and $\Acp(\BtoKstarll)=\AcbBtoKstllHFAG$ as averaged by HFAG.
However, nonzero negative isospin asymmetries in the small
$q^2$-region of $\BtoKll$ ($3.2\sigma$) and $\BtoKstarll$ ($2.7\sigma$)
 have been reported by BaBar
($3.9\sigma$ when combined).  The corresponding 
isospin asymmetries by Belle are $1.4\sigma$ and $1.8\sigma$ from zero, and are
consistent with both BaBar's results and null asymmetry.  The isospin
asymmetry combined for $\BtoKorKstarll$ and averaged by HFAG is
\begin{equation}
  A_I^{K^{(*)}} = \AIBtoKorKstllHFAG.
\end{equation}
The SM prediction is essentially zero at this level of statistics (see
Section~\ref{exclusivepenguins}).

Muon to electron ratios in
$\BtoKorKstarll$ (Eq.~\ref{rk-def}) are also measured by both Belle and
BaBar.  Results are consistent with the SM, and
their na\"{\i}ve averages are $R_K=1.02\pm0.18$ and $R_{K^*}=0.88\pm0.17$.

%

\begin{figure}[ht]
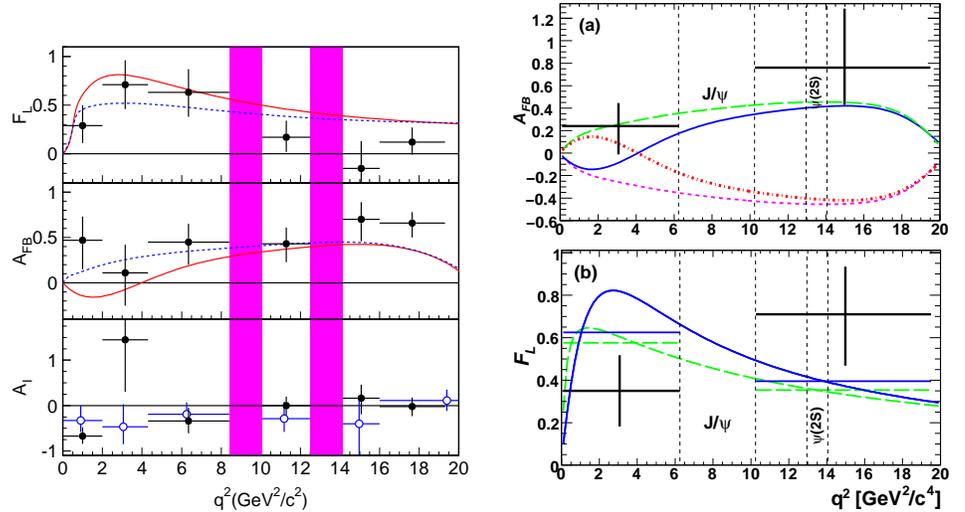

  \begin{minipage}{0.52\textwidth}
    \myeps{belle/kstll/asy_4_3}
  \end{minipage}%
  \begin{minipage}{0.44\textwidth}
    \myeps{babar/afb/fig3a}\\
    \myeps{babar/afb/fig3b}
  \end{minipage}%
  \caption{Longitudinal polarization fraction, forward-backward
    asymmetry and isospin asymmetry of $\BtoKstarll$ by Belle (right,
    from Ref.~\cite{belle-kstll}) and BaBar (left, from
    Ref.~\cite{babar-afbkstll}).  The solid line shows the Standard
    Model predictions, 
    and the other curves represent non-Standard Model extreme cases.}
  \label{fig:both-afb}
\end{figure}

The four-body decay configuration of $\BtoKstarll\to K\pi\elel$ allows
extraction of further information from the angular distributions of
the final-state particles.  The most interesting observables are
the fraction of longitudinal polarization $F_L$ from the kaon angular
distribution (Eq.~\ref{kstarll-thetak}) and the forward-backward asymmetry
$\AFB$ from the lepton angular distribution (Eq.~\ref{kstarll-thetal}).
Belle has measured $\FL$ and $\AFB$ in six bins of
$q^2$~\cite{belle-kstll}, whereas BaBar has done so in two
bins~\cite{babar-afbkstll}.  Current statistics are not enough to tell whether
there is a zero-crossing point at low $q^2$, although the results
favor the case with no crossing, for which the sign of the Wilson
coefficient $C_7$ is flipped.  Both
results have positive $\AFB$ for high $q^2$
(Fig.~\ref{fig:both-afb}), which sets nontrivial
constraints on the Wilson coefficients.  CDF has also measured $\FL$ and
$\AFB$ in the same six bins as Belle for $B^0\to K^{*0}\mumu$ 
events~\cite{cdf-kstll}.  The results are in agreement with Belle and BaBar.

\subsection{Inclusive $\BtoXsll$ Branching Fraction}
\label{sec:ex-xsll}

The inclusive $\BtoXsll$ branching fraction has been measured by Belle
and BaBar using the sum-of-exclusive technique.  The $X_s$ system
includes final states with one kaon and up to four (two) pions that have masses up
to 2.0 (1.8) GeV for the result by Belle (BaBar).  Belle recently
announced a preliminary result based on 657 million
$\BBbar$~\cite{belle-xsll}, and BaBar's result is based on 89
million $\BBbar$~\cite{babar-xsll}.  In Belle's new analysis, partial
branching fractions are measured in bins of the $X_s$ mass,
and then the total branching fraction is calculated as their sum.  This
method reduces the large systematic error observed in previous studies
that arose from the
strong $X_s$ mass dependence of the efficiency and the unknown fractions of
exclusive channels $\BtoKorKstarll$.  The measurement is still dominated
by the statistical error and will be more precise in the future.
Belle and BaBar reported
branching fractions as $\Br(\BtoXsll)=\BrBtoXsllBelle$ and
$\Br(\BtoXsll)=\BrBtoXsllBaBar$, respectively, which were averaged by
HFAG as
\begin{equation}
  \Br(\BtoXsll)=\BrBtoXsllHFAG,
\end{equation}
integrated over the entire subset of phase space with $q^2>0.2\GeV$,
including the vetoed $J/\psi$ and $\psi(2S)$ regions.  The results are
in good agreement with the SM prediction. They  strongly disfavor
the case with the flipped sign of $C_7$~\cite{Gambino:2004mv}.

                       \section{OUTLOOK}
                       \label{sec:outlook}

Remarkably, the $B$ factories have measured  all the
observables within the radiative and electroweak penguin decays at 
values that are consistent with the SM predictions. 
These measurements rule out  $O(1)$  corrections to the SM and identify 
the CKM theory  as the dominant effect for flavor violation as well as
for $CP$ violation.  The success of the simple CKM theory of $CP$ violation was honored with the Nobel 
Prize in Physics in 2008.
Theoretical tools and precision  have  significantly advanced  during
the past decade, and we are ready to challenge 
the SM if  a clear deviation is found or to discriminate  different  NP scenarios if  
direct evidence is found at the LHC.

Also,
the future offers great experimental opportunities in flavor physics.  
LHCb has finally started taking data and promises to overwhelm many
$B$ factory results, and ATLAS and CMS will also contribute to flavor
physics. 
In the radiative and electroweak penguin decays, the most promising
measurements  are  the angular analysis of $B^0\to K^{*0}\mumu$
and the analysis of time-dependent $CP$ asymmetry
in $B_s\to \phi\gamma$; the latter measurement cannot be performed at the $B$ factories due
to the fast $B_s$ oscillation.  However, the theoretically clean inclusive modes 
and many modes involving neutral particles like the $\piZ$   can be
pursued only at the $\epem$ $B$ factories.  Two proposed super-$B$ factories, Belle II at KEK and 
SuperB in Italy,  would accumulate two-orders-of-magnitude-larger
data samples.  Such data would push experimental precision to its  limit.

Theoretical and experimental techniques 
are ready for such  large data samples.  The results
provided by LHCb and the next-generation $\epem$ $B$ factories
are eagerly awaited, as 
they may be the key to identifying physics beyond the SM.
                       

\section*{ACKNOWLEDGMENTS}

We thank Christoph Greub, Colin Jessop,  and Kurtis Nishimura for their careful reading of the
manuscript,
and Thorsten Feldmann, Matthias Neubert,   and Gil Paz for comments.
T.H.\ thanks the CERN Theory Group for its hospitality during his visits to CERN.


\begin{raggedright}

\end{raggedright}


\begin{thebibliography}{99}


\bibitem{Ellis:1977uk}
  Ellis JR, Gaillard MK, Nanopoulos DV, Rudaz S.
  \Journal{\NPB}{131}{285}{1977}, \Journal{Erratum-ibid.}{132}{541}{1978}

\bibitem{Shifman:1995hc}
  Shifman MA.
  arXiv:hep-ph/9510397 (1995)

\bibitem{Belle} 
  Belle collaboration: http://belle.kek.jp/

\bibitem{Babar} 
  BaBar collaboration: http://www.slac.stanford.edu/BFROOT/

\bibitem{Kobayashi:1973fv}
  Kobayashi~M, Maskawa~T. \Journal{\PTP}{49}{652}{1973}

\bibitem{Cabibbo:1963yz}
  Cabibbo~N. \Journal{\PRL}{10}{531}{1963}


\bibitem{TevatronB1}
  CDF collaboration: http://www-cdf.fnal.gov/physics/new/bottom/bottom.html

\bibitem{TevatronB2}
  D0 collaboration: http://www-d0.fnal.gov/Run2Physics/WWW/results/b.htm

\bibitem{Buchalla:2008jp}
  Artuso~M, \etal\ \Journal{\EPJC}{57}{309}{2008}%
  \arXiv{0801.1833}{ [hep-ph]}

\bibitem{Antonelli:2009ws}
  Antonelli~M, \etal\ arXiv:0907.5386 [hep-ph] (2009)

\bibitem{GinoYossi}
  Isidori G, Nir Y, Perez G.
  in the same volume of {\it Ann. Rev. Nucl. Part. Sci.} (2010)%
  \arXiv{1002.0900}{ [hep-ph]}


\bibitem{BurasFlavour} 
  Buras AJ. arXiv:0910.1032 [hep-ph] (2009)


\bibitem{cleo-kstgam1993} \bibBtoKstgamCLEOO
  
\bibitem{cleo-xsgam1995} \bibBtoXsgamICLEOO



\bibitem{Lingel1998ar}
  Lingel K, Skwarnicki T, Smith JG.
  \Journal{Ann. Rev. Nucl. Part. Sci.}{48}{253}{1998}



\bibitem{Hurth:2003vb}
  Hurth T. \Journal{\RMP}{75}{1159}{2003}%
  \arXiv{hep-ph/0212304}{}


\bibitem{Hurth:2003ej}
  Hurth T, Lunghi E.
  {\it In the Proceedings of 2nd Workshop on the CKM Unitarity Triangle,
    Durham, England, 5-9 Apr 2003}
  \arXiv{hep-ph/0307142}{}



\bibitem{Hurth:2007xa}
  Hurth T. \Journal{Int.\ J.\ Mod.\ Phys.\ A}{22}{1781}{2007}%
  \arXiv{hep-ph/0703226}{}



\bibitem{Wilson:1969zs}
  Wilson KG. \Journal{Phys.\ Rev.}{179}{1499}{1969}


\bibitem{Wilson:1970ag}
  Wilson KG. \Journal{\PRD}{3}{1818}{1971}



\bibitem{Gaillard:1974nj}
  Gaillard MK, Lee BW. \Journal{\PRL}{33}{108}{1974}

\bibitem{Altarelli:1974exa}
  Altarelli G, Maiani L. \Journal{\PLB}{52}{351}{1974}

\bibitem{Witten:1976kx}
  Witten E. \Journal{\NPB}{122}{109}{1977}


\bibitem{Chay:1990da}
  Chay J, Georgi H, Grinstein B. \Journal{\PLB}{247}{399}{1990}
 
 \bibitem{Bigi:1992su}
  Bigi II, Uraltsev NG, Vainshtein AI.
  \Journal{\PLB}{293}{430}{1992},
  \Journal{Erratum-ibid}{297}{477}{1993}%
  \arXiv{hep-ph/9207214}{}
  
  \bibitem{Bigi:1992ne}
  Bigi II, \etal\ arXiv:hep-ph/9212227 (1992)
  %
  
 \bibitem{Bigi:1997fj}
  Bigi II, Shifman MA and Uraltsev N.
  \Journal{Ann.\ Rev.\ Nucl.\ Part.\ Sci.}{47}{591}{1997}%
  \arXiv{hep-ph/9703290}{}
  
 \bibitem{Manohar:1993qn}
  Manohar AV and Wise MB. \Journal{\PRD}{49}{1310}{1994}%
  \arXiv{hep-ph/9308246}{}

 \bibitem{Manohar:2000dt}
  Manohar AV, Wise MB.
  \Journal{Camb.\ Monogr.\ Part.\ Phys.\ Nucl.\ Phys.\ Cosmol.}{10}{1}{2000}
  

\bibitem{Falk:1993dh}
  Falk AF, Luke ME, Savage MJ. \Journal{\PRD}{49}{3367}{1994}%
  \arXiv{hep-ph/9308288}{}



\bibitem{Ali:1996bm}
  Ali A, Hiller G, Handoko LT, Morozumi T. \Journal{\PRD}{55}{4105}{1997}%
  \arXiv{hep-ph/9609449}{}



\bibitem{Beneke:1999br}
  Beneke M, Buchalla G, Neubert M, Sachrajda CT. \Journal{\PRL}{83}{1914}{1999}%
  \arXiv{hep-ph/9905312}{}

\bibitem{Beneke:2000ry}
  Beneke M, Buchalla G, Neubert M, Sachrajda CT. \Journal{\NPB}{591}{313}{2000}%
  \arXiv{hep-ph/0006124}{}

\bibitem{Beneke:2001ev}
  Beneke M, Buchalla G, Neubert M, Sachrajda CT. \Journal{\NPB}{606}{245}{2001}%
  \arXiv{hep-ph/0104110}{}



\bibitem{Bauer:2000ew}
  Bauer CW, Fleming S, Luke ME. \Journal{\PRD}{63}{014006}{2000}%
  \arXiv{hep-ph/0005275}{}

\bibitem{Bauer:2000yr}
  Bauer CW, Fleming S, Pirjol D, Stewart IW. \Journal{\PRD}{63}{114020}{2001}%
  \arXiv{hep-ph/0011336}{}

\bibitem{Bauer:2001ct}
  Bauer CW, Stewart IW. \Journal{\PLB}{516}{134}{2001}%
  \arXiv{hep-ph/0107001}{}

\bibitem{Bauer:2001yt}
  Bauer CW, Pirjol D, Stewart IW. \Journal{\PRD}{65}{054022}{2002}%
  \arXiv{hep-ph/0109045}{}


\bibitem{Beneke:2002ph}
  Beneke M, Chapovsky AP, Diehl M, Feldmann T. \Journal{\NPB}{643}{431}{2002}%
  \arXiv{hep-ph/0206152}{}

\bibitem{Hill:2002vw}
  Hill RJ, Neubert M. \Journal{\NPB}{657}{229}{2003}%
  \arXiv{hep-ph/0211018}{}




\bibitem{Ciuchini:1993ks}
  Ciuchini M \etal\ \Journal{\PLB}{316}{127}{1993}%
  \arXiv{hep-ph/9307364}{}

\bibitem{Ciuchini:1993vr}
  Ciuchini M, Franco E, Martinelli G, Reina L. \Journal{\NPB}{415}{403}{1994}%
  \arXiv{hep-ph/9304257}{}

\bibitem{Cella:1994np}
  Cella G, Curci G, Ricciardi G, Vicere A. \Journal{\NPB}{431}{417}{1994}%
  \arXiv{hep-ph/9406203}{}

\bibitem{Misiak:1992bc}
  Misiak M.
  \Journal{\NPB}{393}{23}{1993},
  \Journal{Erratum-ibid}{439}{461}{1995}



\bibitem{Adel:1993ah}
  Adel K, Yao YP. \Journal{\PRD}{49}{4945}{1994}%
  \arXiv{hep-ph/9308349}{}

\bibitem{Greub:1997hf}
  Greub C, Hurth T. \Journal{\PRD}{56}{2934}{1997}%
  \arXiv{hep-ph/9703349}{}


\bibitem{Chetyrkin:1996vx}
  Chetyrkin KG, Misiak M, Munz M.
  \Journal{\PLB}{400}{206}{1997},
  \Journal{Erratum-ibid}{425}{414}{1998}%
  \arXiv{hep-ph/9612313}{}

\bibitem{Gambino:2003zm}
  Gambino P, Gorbahn M, Haisch U. \Journal{\NPB}{673}{238}{2003}%
  \arXiv{hep-ph/0306079}{}



\bibitem{Ali:1990tj}
  Ali A, Greub C. \Journal{\ZPC}{49}{431}{1991}



\bibitem{Greub:1996tg}
  Greub C, Hurth T, Wyler D. \Journal{\PRD}{54}{3350}{1996}%
  \arXiv{hep-ph/9603404}{}



\bibitem{Pott:1995if}
  Pott N. \Journal{\PRD}{54}{938}{1996}%
  \arXiv{hep-ph/9512252}{}

\bibitem{Buras:2001mq}
  Buras AJ, Czarnecki A, Misiak M, Urban J. \Journal{\NPB}{611}{488}{2001}%
  \arXiv{hep-ph/0105160}{}

\bibitem{Buras:2002tp}
  Buras AJ, Czarnecki A, Misiak M, Urban J. \Journal{\NPB}{631}{219}{2002}%
  \arXiv{hep-ph/0203135}{}



\bibitem{Gambino:2001ew}
  Gambino P, Misiak M. \Journal{\NPB}{611}{338}{2001}%
  \arXiv{hep-ph/0104034}{}

\bibitem{Hurth:2003dk}
  Hurth T, Lunghi E, Porod W. \Journal{\NPB}{704}{56}{2005}%
  \arXiv{hep-ph/0312260}{}
 
\bibitem{Hurth:2003pn}
  Hurth T, Lunghi E, Porod W. \Journal{\EPJC}{33}{s382}{2004}%
  \arXiv{hep-ph/0310282}{}

\bibitem{Asatrian:2005pm}
  Asatrian HM \etal\ \Journal{\PLB}{619}{322}{2005}%
  \arXiv{hep-ph/0505068}{}
  %
  
  

\bibitem{Misiak:2006zs}
  Misiak M, \etal\ \Journal{\PRL}{98}{022002}{2007}%
  \arXiv{hep-ph/0609232}{}

\bibitem{Misiak:2004ew}
  Misiak M, Steinhauser M. \Journal{\NPB}{683}{277}{2004}%
  \arXiv{hep-ph/0401041}{}

\bibitem{Bobeth:1999mk}
  Bobeth C, Misiak M, Urban J. \Journal{\NPB}{574}{291}{2000}%
  \arXiv{hep-ph/9910220}{}


\bibitem{Gorbahn:2004my}
  Gorbahn M, Haisch U. \Journal{\NPB}{713}{291}{2005}%
  \arXiv{hep-ph/0411071}{}

\bibitem{Gorbahn:2005sa}
  Gorbahn M, Haisch U, Misiak M. \Journal{\PRL}{95}{102004}{2005}%
  \arXiv{hep-ph/0504194}{}

\bibitem{Czakon:2006ss}
  Czakon M, Haisch U, Misiak M. \Journal{JHEP}{0703}{008}{2007}%
  \arXiv{hep-ph/0612329}{}

\bibitem{Melnikov:2005bx}
  Melnikov K, Mitov A. \Journal{\PLB}{620}{69}{2005}%
  \arXiv{hep-ph/0505097}{}

\bibitem{Blokland:2005uk}
  Blokland I \etal\ \Journal{\PRD}{72}{033014}{2005}%
  \arXiv{hep-ph/0506055}{}
  %

\bibitem{Asatrian:2006ph}
  Asatrian HM, \etal\ \Journal{\NPB}{749}{325}{2006}%
  \arXiv{hep-ph/0605009}{}
  %

\bibitem{Asatrian:2006sm}
  Asatrian HM, \etal\ \Journal{\NPB}{762}{212}{2007}%
  \arXiv{hep-ph/0607316}{}

\bibitem{Bieri:2003ue}
  Bieri K, Greub C, Steinhauser M. \Journal{\PRD}{67}{114019}{2003}%
  \arXiv{hep-ph/0302051}{}

\bibitem{Misiak:2006ab}
  Misiak M, Steinhauser M. \Journal{\NPB}{764}{62}{2007}%
  \arXiv{hep-ph/0609241}{}

\bibitem{Boughezal:2007ny}
  Boughezal R, Czakon M, Schutzmeier T, \Journal{JHEP}{0709}{072}{2007}%
  \arXiv{0707.3090}{ [hep-ph]}

\bibitem{Ligeti:1999ea}
  Ligeti Z, Luke ME, Manohar AV, Wise MB. \Journal{\PRD}{60}{034019}{1999}%
  \arXiv{hep-ph/9903305}{}

\bibitem{Asatrian:2006rq}
  Asatrian HM, Ewerth T, Gabrielyan H, Greub C. \Journal{\PLB}{647}{173}{2007}%
  \arXiv{hep-ph/0611123}{}

\bibitem{Ewerth:2008nv}
  Ewerth T. \Journal{\PLB}{669}{167}{2008}%
  \arXiv{0805.3911}{ [hep-ph]}


  
\bibitem{Czarnecki:1998tn}
  Czarnecki A, Marciano WJ. \Journal{\PRL}{81}{277}{1998}%
  \arXiv{hep-ph/9804252}{}

\bibitem{Kagan:1998ym}
  Kagan AL, Neubert M. \Journal{\EPJC}{7}{5}{1999}%
  \arXiv{hep-ph/9805303}{}

\bibitem{Baranowski:1999tq}
  Baranowski K, Misiak M. \Journal{\PLB}{483}{410}{2000}%
  \arXiv{hep-ph/9907427}{}

\bibitem{Gambino:2000fz}
  Gambino P, Haisch U. \Journal{JHEP}{0110}{020}{2001}%
  \arXiv{hep-ph/0109058}{}

  


\bibitem{Buras:1994dj}
  Buras AJ, Munz M. \Journal{\PRD}{52}{186}{1995}%
  \arXiv{hep-ph/9501281}{}


\bibitem{Asatryan:2001zw}
  Asatryan HH, Asatrian HM, Greub C, Walker M. \Journal{\PRD}{65}{074004}{2002}%
  \arXiv{hep-ph/0109140}{}

\bibitem{Asatryan:2002iy}
  Asatryan HH, Asatrian HM, Greub C, Walker M. \Journal{\PRD}{66}{034009}{2002}%
  \arXiv{hep-ph/0204341}{}

\bibitem{Ghinculov:2003qd}
  Ghinculov A, Hurth T, Isidori G, Yao YP. \Journal{\NPB}{685}{351}{2004}%
  \arXiv{hep-ph/0312128}{}

\bibitem{Ghinculov:2003bx}
  Ghinculov A, Hurth T, Isidori G, Yao YP. \Journal{\EPJC}{33}{s288}{2004}%
  \arXiv{hep-ph/0310187}{}


\bibitem{Asatrian:2002va}
  Asatrian HM, Bieri K, Greub C, Hovhannisyan A.
  \Journal{\PRD}{66}{094013}{2002}%
  \arXiv{hep-ph/0209006}{}

\bibitem{Asatrian:2003yk}
  Asatrian HM, Asatryan HH, Hovhannisyan A, Poghosyan V.
  \Journal{Mod.\ Phys.\ Lett.\ A}{19}{603}{2004}%
  \arXiv{hep-ph/0311187}{}

\bibitem{Ghinculov:2002pe}
  Ghinculov A, Hurth T, Isidori G, Yao YP. \Journal{\NPB}{648}{254}{2003}%
  \arXiv{hep-ph/0208088}{}


\bibitem{Greub:2008cy}
  Greub C, Pilipp V, Schupbach C. \Journal{JHEP}{0812}{040}{2008}%
  \arXiv{0810.4077}{ [hep-ph]}

\bibitem{Huber:2008ak}
  Huber T, Hurth T, Lunghi E. arXiv:0807.1940 [hep-ph] (2008)



\bibitem{Bobeth:2003at}
  Bobeth C, Gambino P, Gorbahn M, Haisch U. \Journal{JHEP}{0404}{071}{2004}%
  \arXiv{hep-ph/0312090}{}

\bibitem{Huber:2005ig}
  Huber T, Lunghi E, Misiak M, Wyler D. \Journal{\NPB}{740}{105}{2006}%
  \arXiv{hep-ph/0512066}{}

\bibitem{Huber:2007vv}
  Huber T, Hurth T, Lunghi E. \Journal{\NPB}{802}{40}{2008}%
  \arXiv{0712.3009}{ [hep-ph]}


\bibitem{Ali:1998rr}
  Ali A, Asatrian H, Greub C. \Journal{\PLB}{429}{87}{1998}%
  \arXiv{hep-ph/9803314}{}




\bibitem{Asatrian:2003vq}
  Asatrian HM, Bieri K, Greub C, Walker M. \Journal{\PRD}{69}{074007}{2004}%
  \arXiv{hep-ph/0312063}{}


\bibitem{Seidel:2004jh}
  Seidel D. \Journal{\PRD}{70}{094038}{2004}%
  \arXiv{hep-ph/0403185}{}



\bibitem{Isgur:1991xa}
  Isgur N, Wise MB.
  \Journal{Adv.\ Ser.\ Direct.\ High Energy Phys.}{10}{549}{1992}
  
\bibitem{Neubert:1993mb}
  Neubert M. \Journal{Phys.\ Rept.}{245}{259}{1994}%
  \arXiv{hep-ph/9306320}{}
  

\bibitem{Ligeti:1997tc}
  Ligeti Z, Randall L, Wise MB. \Journal{\PLB}{402}{178}{1997}%
  \arXiv{hep-ph/9702322}{}

\bibitem{Voloshin:1996gw}
  Voloshin MB. \Journal{\PLB}{397}{275}{1997}%
  \arXiv{hep-ph/9612483}{}

\bibitem{Benzke:2010js}
  Benzke M, Lee SJ, Neubert M, Paz G. arXiv:1003.5012 [hep-ph] (2010)

\bibitem{Grant:1997ec}
  Grant AK, Morgan AG, Nussinov S, Peccei RD. \Journal{\PRD}{56}{3151}{1997}%
  \arXiv{hep-ph/9702380}{}

\bibitem{Buchalla:1997ky}
  Buchalla G, Isidori G, Rey SJ. \Journal{\NPB}{511}{594}{1998}%
  \arXiv{hep-ph/9705253}{}
 
\bibitem{Lee:2006wn}
  Lee SJ, Neubert M, Paz G. \Journal{\PRD}{75}{114005}{2007}%
  \arXiv{hep-ph/0609224}{}
  
\bibitem{Lee:2004ja}
  Lee KSM, Stewart IW. \Journal{\NPB}{721}{325}{2005}%
  \arXiv{hep-ph/0409045}{}

\bibitem{Bosch:2004cb}
  Bosch SW, Neubert M, Paz G. \Journal{JHEP}{0411}{073}{2004}%
  \arXiv{hep-ph/0409115}{}

\bibitem{Beneke:2004in}
  Beneke M, Campanario F, Mannel T, Pecjak BD. \Journal{JHEP}{0506}{071}{2005}%
  \arXiv{hep-ph/0411395}{}

  
\bibitem{Chen:1997dj}
  Chen JW, Rupak G, Savage MJ. \Journal{\PLB}{410}{285}{1997}%
  \arXiv{hep-ph/9705219}{}

\bibitem{Buchalla:1998mt}
  Buchalla G, Isidori G. \Journal{\NPB}{525}{333}{1998}%
  \arXiv{hep-ph/9801456}{}


\bibitem{Bauer:1999kf}
  Bauer CW, Burrell CN. \Journal{\PRD}{62}{114028}{2000}%
  \arXiv{hep-ph/9911404}{}

\bibitem{Ligeti:2007sn}
  Ligeti Z, Tackmann FJ. \Journal{\PLB}{653}{404}{2007}%
  \arXiv{0707.1694}{ [hep-ph]}

\bibitem{Neubert:2000ch}
  Neubert M. \Journal{JHEP}{0007}{022}{2000}%
  \arXiv{hep-ph/0006068}{}

\bibitem{Bauer:2001rc}
  Bauer CW, Ligeti Z, Luke ME. \Journal{\PRD}{64}{113004}{2001}%
  \arXiv{hep-ph/0107074}{}











\bibitem{Kruger:1996cv}
  Kruger F, Sehgal LM. \Journal{\PLB}{380}{199}{1996}%
  \arXiv{hep-ph/9603237}{}

\bibitem{Kruger:1996dt}
  Kruger F, Sehgal LM. \Journal{\PRD}{55}{2799}{1997}%
  \arXiv{hep-ph/9608361}{}




%
%
%

\bibitem{Beneke:2009az}
  Beneke M, Buchalla G, Neubert M, Sachrajda CT. \Journal{\EPJC}{61}{439}{2009}%
  \arXiv{0902.4446}{ [hep-ph]}

\bibitem{Benson:2004sg}
  Benson D, Bigi II, Uraltsev NG. \Journal{\NPB}{710}{371}{2005}%
  \arXiv{hep-ph/0410080}{}

\bibitem{Bosch:2004th}
  Bosch SW, Lange BO, Neubert M, Paz G. \Journal{\NPB}{699}{335}{2004}%
  \arXiv{hep-ph/0402094}{}

\bibitem{Buchmuller:2005zv}
  Buchmuller O, Flacher H. \Journal{\PRD}{73}{073008}{2006}%
  \arXiv{hep-ph/0507253}{}

\bibitem{Neubert:2004dd}
  Neubert M. \Journal{\EPJC}{40}{165}{2005}%
  \arXiv{hep-ph/0408179}{}

\bibitem{Becher:2006pu}
  Becher T, Neubert M. \Journal{\PRL}{98}{022003}{2007}%
  \arXiv{hep-ph/0610067}{}

\bibitem{Becher:2005pd}
  Becher T, Neubert M. \Journal{\PLB}{633}{739}{2006}%
  \arXiv{hep-ph/0512208}{}

\bibitem{Becher:2006qw}
  Becher T, Neubert M. \Journal{\PLB}{637}{251}{2006}%
  \arXiv{hep-ph/0603140}{}

\bibitem{Misiak:2008ss}
  Misiak M. arXiv:0808.3134 [hep-ph] (2008)


\bibitem{Andersen:2006hr}
  Andersen JR, Gardi E. \Journal{JHEP}{0701}{029}{2007}%
  \arXiv{hep-ph/0609250}{}

\bibitem{Andersen:2005bj}
  Andersen JR, Gardi E. \Journal{JHEP}{0506}{030}{2005}%
  \arXiv{hep-ph/0502159}{}

\bibitem{Gardi:2006jc}
  Gardi E. arXiv:hep-ph/0606080 (2006)



\bibitem{Lee:2005pk}
  Lee KSM, Stewart IW. \Journal{\PRD}{74}{014005}{2006}%
  \arXiv{hep-ph/0511334}{}

\bibitem{Lee:2005pw}
  Lee KSM, Ligeti Z, Stewart IW, Tackmann FJ.
  \Journal{\PRD}{74}{011501}{2006}%
  \arXiv{hep-ph/0512191}{}

\bibitem{Lee:2008xc}
  Lee KSM, Tackmann FJ. \Journal{\PRD}{79}{114021}{2009}%
  \arXiv{0812.0001}{ [hep-ph]}



\bibitem{Poggio:1975af}
  Poggio EC, Quinn HR, Weinberg S. \Journal{\PRD}{13}{1958}{1976}



\bibitem{Beneke:2000wa}
  Beneke M, Feldmann T. \Journal{\NPB}{592}{3}{2001}%
  \arXiv{hep-ph/0008255}{}

\bibitem{Beneke:2001at}
  Beneke M, Feldmann T, Seidel D. \Journal{\NPB}{612}{25}{2001}%
  \arXiv{hep-ph/0106067}{}

\bibitem{Bosch:2001gv}
  Bosch SW, Buchalla G. \Journal{\NPB}{621}{459}{2002}%
  \arXiv{hep-ph/0106081}{}

\bibitem{Ali:2001ez}
  Ali A, Parkhomenko AY. \Journal{\EPJC}{23}{89}{2002}%
  \arXiv{hep-ph/0105302}{}

\bibitem{Descotes-Genon:2004hd}
  Descotes-Genon S, Sachrajda CT. \Journal{\NPB}{693}{103}{2004}%
  \arXiv{hep-ph/0403277}{}


\bibitem{Ali:2007sj}
  Ali A, Pecjak BD and C.~Greub C. \Journal{\EPJC}{55}{577}{2008}%
  \arXiv{0709.4422}{ [hep-ph]}




\bibitem{Braun:1988qv}
  Braun VM, Filyanov IE. \Journal{\ZPC}{44}{157}{1989}

\bibitem{Braun:1989iv}
  Braun VM, Filyanov IE. \Journal{\ZPC}{48}{239}{1990}

\bibitem{Ball:1998sk}
  Ball P, Braun VM, Koike Y, Tanaka K. \Journal{\NPB}{529}{323}{1998}%
  \arXiv{hep-ph/9802299}{}

\bibitem{Ball:1998ff}
  Ball P, Braun VM. \Journal{\NPB}{543}{201}{1999}%
  \arXiv{hep-ph/9810475}{}



\bibitem{Bauer:2002aj}
  Bauer CW, Pirjol D, Stewart IW. \Journal{\PRD}{67}{071502}{2003}%
  \arXiv{hep-ph/0211069}{}

\bibitem{Beneke:2003pa}
  Beneke M, Feldmann T. \Journal{\NPB}{685}{249}{2004}%
  \arXiv{hep-ph/0311335}{}

\bibitem{Lange:2003pk}
  Lange BO, Neubert M. \Journal{\NPB}{690}{249}{2004}%
  \arXiv{hep-ph/0311345}{}

\bibitem{Charles:1998dr}
  Charles J, \etal\ \Journal{\PRD}{60}{014001}{1999}%
  \arXiv{hep-ph/9812358}{}

\bibitem{Becher:2005fg}
  Becher T, Hill RJ, Neubert M. \Journal{\PRD}{72}{094017}{2005}%
  \arXiv{hep-ph/0503263}{}

\bibitem{Feldmann:2004mg}
  Feldmann T, Hurth T. \Journal{JHEP}{0411}{037}{2004}%
  \arXiv{hep-ph/0408188}{}

\bibitem{Kagan:2001zk}
  Kagan AL, Neubert M. \Journal{\PLB}{539}{227}{2002}%
  \arXiv{hep-ph/0110078}{}

\bibitem{Becher:2003qh}
  Becher T, Hill RJ, Neubert M. \Journal{\PRD}{69}{054017}{2004}%
  \arXiv{hep-ph/0308122}{}

\bibitem{Arnesen:2006vb}
  Arnesen CM, Ligeti Z, Rothstein IZ, Stewart IW.
  \Journal{\PRD}{77}{054006}{2008}%
  \arXiv{hep-ph/0607001}{}


\bibitem{Ali:2004hn}
  Ali A, Lunghi E, Parkhomenko AY. \Journal{\PLB}{595}{323}{2004}%
  \arXiv{hep-ph/0405075}{}

\bibitem{Bosch:2004nd}
  Bosch SW, Buchalla G. \Journal{JHEP}{0501}{035}{2005}%
  \arXiv{hep-ph/0408231}{}

\bibitem{Beneke:2004dp}
  Beneke M, Feldmann T, Seidel D. \Journal{\EPJC}{41}{173}{2005}%
  \arXiv{hep-ph/0412400}{}


\bibitem{Ball:2006eu}
  Ball P, Jones GW, Zwicky R. \Journal{\PRD}{75}{054004}{2007}%
  \arXiv{hep-ph/0612081}{}



\bibitem{Egede:2008uy}
  Egede U, \etal\ \Journal{JHEP}{0811}{032}{2008}%
  \arXiv{0807.2589}{ [hep-ph]}



\bibitem{Bobeth:2007dw}
  Bobeth C, Hiller G, Piranishvili G. \Journal{JHEP}{0712}{040}{2007}%
  \arXiv{0709.4174}{ [hep-ph]}



\bibitem{babar-kstgam}   \bibBtoKstgamBaBar\arxivBtoKstgamBaBar

\bibitem{pythia}
  Sjostrand T, Mrenna S, Skands P. \Journal{JHEP}{0605}{026}{2006}%
  \arXiv{hep-ph/0603175}{}



\bibitem{Gambino:2008fj}
  Gambino P, Giordano P. \Journal{\PLB}{669}{69}{2008}%
  \arXiv{0805.0271}{ [hep-ph]}


\bibitem{Ciuchini:1997xe}
  Ciuchini M, Degrassi G, Gambino P, Giudice GF.
  \Journal{\NPB}{527}{21}{1998}%
  \arXiv{hep-ph/9710335}{}

\bibitem{Borzumati:1998tg}
  Borzumati F, Greub C. \Journal{\PRD}{58}{074004}{1998}%
  \arXiv{hep-ph/9802391}{}

\bibitem{Haisch:2007vb}
  Haisch U, Weiler A. \Journal{\PRD}{76}{034014}{2007}%
  \arXiv{hep-ph/0703064}{}

\bibitem{Bertolini:1990if}
  Bertolini S, Borzumati F, Masiero A, Ridolfi G.
  \Journal{\NPB}{353}{591}{1991}

\bibitem{Degrassi:2000qf}
  Degrassi G, Gambino P, Giudice GF. \Journal{JHEP}{0012}{009}{2000}%
  \arXiv{hep-ph/0009337}{}

\bibitem{Carena:2000uj}
  Carena MS, Garcia D, Nierste U, Wagner CEM.
  \Journal{\PLB}{499}{141}{2001}%
  \arXiv{hep-ph/0010003}{}

\bibitem{Degrassi:2006eh}
  Degrassi G, Gambino P, Slavich P. \Journal{\PLB}{635}{335}{2006}%
  \arXiv{hep-ph/0601135}{}

\bibitem{Borzumati:1999qt}
  Borzumati F, Greub C, Hurth T, Wyler D. \Journal{\PRD}{62}{075005}{2000}%
  \arXiv{hep-ph/9911245}{}

\bibitem{Besmer:2001cj}
  Besmer T, Greub C, Hurth T. \Journal{\NPB}{609}{359}{2001}%
  \arXiv{hep-ph/0105292}{}

\bibitem{Ciuchini:2002uv}
  Ciuchini M, Franco E, Masiero A, Silvestrini L.
  \Journal{\PRD}{67}{075016}{2003},
  \Journal{Erratum-ibid}{68}{079901}{2003}%
  \arXiv{hep-ph/0212397}{}

\bibitem{Ciuchini:2003rg}
  Ciuchini M, \etal\ \Journal{\PRL}{92}{071801}{2004}%
  \arXiv{hep-ph/0307191}{}
  %

\bibitem{Altmannshofer:2008vr}
  Altmannshofer W, Guadagnoli D, Raby S, Straub DM.
  \Journal{\PLB}{668}{385}{2008}%
  \arXiv{0801.4363}{ [hep-ph]}




\bibitem{Altmannshofer:2009ne}
  Altmannshofer W, \etal\ \Journal{\NPB}{830}{17}{2010}%
  \arXiv{0909.1333}{ [hep-ph]}
  %

\bibitem{Blanke:2009am}
  Blanke M, \etal\ \Journal{Acta Phys. Polon. B}{41}{657}{2010}%
  \arXiv{0906.5454}{ [hep-ph]}

\bibitem{Ali:2002jg}
  Ali A, Lunghi E, Greub C, Hiller G. \Journal{\PRD}{66}{034002}{2002}%
  \arXiv{hep-ph/0112300}{}

\bibitem{D'Ambrosio:2002ex}
  D'Ambrosio G, Giudice GF, Isidori G, Strumia A.
  \Journal{\NPB}{645}{155}{2002}%
  \arXiv{hep-ph/0207036}{}

\bibitem{Hurth:2008jc}
  Hurth T, Isidori G, Kamenik JF, Mescia F. \Journal{\NPB}{808}{326}{2009}%
  \arXiv{0807.5039}{ [hep-ph]}



\bibitem{Kagan:1998bh}
  Kagan AL, Neubert M. \Journal{\PRD}{58}{094012}{1998}%
  \arXiv{hep-ph/9803368}{}

\bibitem{Soares:1991te}
  Soares JM. \Journal{\NPB}{367}{575}{1991}

\bibitem{Hurth:2001yb}
  Hurth T, Mannel T. \Journal{\PLB}{511}{196}{2001}%
  \arXiv{hep-ph/0103331}{}

\bibitem{Hurth:2001ja}
  Hurth T, Mannel T. \Journal{AIP Conf.\ Proc.}{602}{212}{2001}%
  \arXiv{hep-ph/0109041}{}


\bibitem{Lee:2006gs}
  Lee KSM, Ligeti Z, Stewart IW, Tackmann FJ.
  \Journal{\PRD}{75}{034016}{2007}%
  \arXiv{hep-ph/0612156}{}

\bibitem{Gambino:2004mv}
  Gambino P, Haisch U, Misiak M. \Journal{\PRL}{94}{061803}{2005}%
  \arXiv{hep-ph/0410155}{}


\bibitem{yan-rkkll}     \bibRKBtoKllYan\arXiv{hep-ph/0004262}{}
\bibitem{hiller-rkkll}  \bibRKBtoKllHiller\arXiv{hep-ph/0310219}{}



\bibitem{Ball:2006nr}
  Ball P, Zwicky R. \Journal{JHEP}{0604}{046}{2006}%
  \arXiv{hep-ph/0603232}{}


\bibitem{Grinstein:2004uu}
  Grinstein B, Grossman Y, Ligeti Z, Pirjol D. \Journal{\PRD}{71}{011504}{2005}%
  \arXiv{hep-ph/0412019}{}


\bibitem{Grinstein:2005nu}
  Grinstein B, Pirjol D. \Journal{\PRD}{73}{014013}{2006}%
  \arXiv{hep-ph/0510104}{}

\bibitem{Ball:2006cva}
  Ball P, Zwicky R. \Journal{\PLB}{642}{478}{2006}%
  \arXiv{hep-ph/0609037}{}


\bibitem{Feldmann:2002iw}
  Feldmann T, Matias J. \Journal{JHEP}{0301}{074}{2003}%
  \arXiv{hep-ph/0212158}{}

\bibitem{Egedeneu}
  Egede U, \etal\  arXiv:1005.0571 [hep-ph] (2010)

\bibitem{Kruger:1999xa}
  Kruger F, Sehgal LM, Sinha N, Sinha R.
  \Journal{\PRD}{61}{114028}{2000}
  \Journal{Erratum-ibid.}{63}{019901}{2001}%
  \arXiv{hep-ph/9907386}{}


\bibitem{Kruger:2005ep}
  Kruger F, Matias J. \Journal{\PRD}{71}{094009}{2005}%
  \arXiv{hep-ph/0502060}{}



\bibitem{Altmannshofer:2008dz}
  Altmannshofer W, \etal\ \Journal{JHEP}{0901}{019}{2009}%
  \arXiv{0811.1214}{ [hep-ph]}


\bibitem{Bobeth:2008ij}
  Bobeth C, Hiller G, Piranishvili G. \Journal{JHEP}{0807}{106}{2008}%
  \arXiv{0805.2525}{ [hep-ph]}

\bibitem{Egede:2009tp}
  Egede U, \etal\ arXiv:0912.1349 [hep-ph] (2009)



\bibitem{cleo-xsgami} \bibBtoXsgamICLEO\arxivBtoXsgamICLEO
\bibitem{babar-xsgami} \bibBtoXsgamIBaBar\arxivBtoXsgamIBaBar
\bibitem{belle-xsgamo} \bibBtoXsgamIBelleO\arxivBtoXsgamIBelleO
\bibitem{belle-xsgami} \bibBtoXsgamIBelle\arxivBtoXsgamIBelle
\bibitem{babar-xsgamx} \bibBtoXsgamXBaBar\arxivBtoXsgamXBaBar
\bibitem{babar-xsgamb} \bibBtoXsgamBBaBar\arxivBtoXsgamBBaBar
\bibitem{belle-xsgamx} \bibBtoXsgamXBelle\arxivBtoXsgamXBelle

\bibitem{hfag}
Barberio E, \etal\ 
arXiv:0808.1297 and online update at\\
http://www.slac.stanford.edu/xorg/hfag (2009)



\bibitem{belle-kstgam}   \bibBtoKstgamBelle \arxivBtoKstgamBelle


\bibitem{belle-kxgam}    \bibBtoKXgamBelle\arxivBtoKXgamBelle
\bibitem{babar-k2stgam}  \bibBtoKtwogamBaBar\arxivBtoKtwogamBaBar
\bibitem{belle-k1gam}    \bibBtoKonegamBelle\arxivBtoKonegamBelle
\bibitem{babar-kpipigam} \bibBtoKpipigamBaBar\arxivBtoKpipigamBaBar
\bibitem{belle-kphigam}  \bibBtoKphigamBelle\arxivBtoKphigamBelle
\bibitem{babar-kphigam}  \bibBtoKphigamBaBar\arxivBtoKphigamBaBar
\bibitem{belle-ketagam}  \bibBtoKetagamBelle\arxivBtoKetagamBelle
\bibitem{babar-ketagam}  \bibBtoKetagamBaBar\arxivBtoKetagamBaBar
\bibitem{belle-ketapgam} \bibBtoKetapgamBelle
\bibitem{belle-pLgam}    \bibBtopLgamBelle\arxivBtopLgamBelle

\bibitem{belle-acpxsgam} \bibAcpBtoXsgamXBelle\arxivAcpBtoXsgamXBelle
\bibitem{babar-acpxsgam} \bibAcpBtoXsgamXBaBar\arxivAcpBtoXsgamXBaBar

\bibitem{pdg}  
Amsler C, et al. (Particle Data Group) \Journal{\PLB}{667}{1}{2008}
\bibitem{atwood-tcpvpqgam}   \bibTCPVBtoPQgamAtwood\arXiv{hep-ph/0410036}{}

\bibitem{belle-tcpvkspizgam} \bibTCPVBtoKSpiZgamBelle\arxivTCPVBtoKSpiZgamBelle
\bibitem{babar-tcpvkspizgam} \bibTCPVBtoKSpiZgamBaBar\arxivTCPVBtoKSpiZgamBaBar

\bibitem{belle-tcpvksrhogam} \bibTCPVBtoKSrhogamBelle\arxivTCPVBtoKSrhogamBelle

\bibitem{belle-rhogam}  \bibBtorhogamBelle\arxivBtorhogamBelle
\bibitem{babar-rhogam}  \bibBtorhogamBaBar\arxivBtorhogamBaBar

\bibitem{babar-xdgam}       \bibBtoXdgamBaBar\arxivBtoXdgamBaBar
\bibitem{belle-tcpvrhogam}  \bibTCPVBtorhogamBelle\arxivTCPVBtorhogamBelle



\bibitem{belle-kstll}    \bibBtoKstllBelle\arxivBtoKstllBelle
\bibitem{babar-kstll}    \bibBtoKstllBaBar\arxivBtoKstllBaBar

\bibitem{cdf-kstll}   CDF Public note 10047 (2009) 



\bibitem{babar-afbkstll} \bibAFBBtoKstllBaBar\arxivAFBBtoKstllBaBar

\bibitem{belle-xsll} 
  In talk given by T.~Iijima at {\it XXIV International Symposium on
    Lepton Photon Interactions}\/ (2009)


\bibitem{babar-xsll} \bibBtoXsllBaBar\arxivBtoXsllBaBar

\end{thebibliography}
\end{document}